\documentclass[useAMS, usenatbib]{mnras}

\usepackage{newtxtext,newtxmath}

\usepackage[T1]{fontenc}
\usepackage{ae,aecompl}

\usepackage{graphicx}	

\usepackage{epsfig}
\usepackage{subfigure}
\usepackage{multirow}

\def \aj{Astronom.~J.}

\def \aap{A\&A }
\def \apjl{ApJ}
\def \apjs{ApJS}
\def \apj{ApJ}

\def \mnras{MNRAS}
\def \prl{Phys.~Rev.~Lett.,}
\def \prd{Phys.~Rev.~D,}

\def \nat{Nature\ }

\newcommand{\msun}{M_{\odot}}

\newcommand{\fg}{f_\mathrm{\rm gas}}

\newcommand{\MA}{\mathcal{M}_{\rm A}}

\newcommand{\uLA}{u_{\mathrm{LA}}}

\newcommand{\vAi}{v_{\mathrm{A,i}}}

\newcommand{\nH}{n_{\mathrm{H}}}
\newcommand{\mH}{m_{\mathrm{H}}}
\newcommand{\muH}{\mu_{\mathrm{H}}}
\newcommand{\gd}{\gamma_{\mathrm{d}}}
\newcommand{\nHthree}{n_{\mathrm{H},3}}
\newcommand{\uLAone}{u_{\mathrm{LA,1}}}

\newcommand{\mui}{\mu_{\mathrm{i}}}

\newcommand{\ECRz}{E_{\mathrm{CR,0}}}

\newcommand{\nCR}{n_{\mathrm{CR}}}
\newcommand{\vst}{v_s}

\newcommand{\appropto}{\mathrel{\vcenter{
  \offinterlineskip\halign{\hfil$##$\cr
    \propto\cr\noalign{\kern2pt}\sim\cr\noalign{\kern-2pt}}}}}

\def\alt{\raise0.3ex\hbox{$\;<$\kern-0.75em\raise-1.1ex\hbox{$\sim\;$}}}
\def\agt{\raise0.3ex\hbox{$\;>$\kern-0.75em\raise-1.1ex\hbox{$\sim\;$}}}

\newcommand{\bw}{\begin{widetext}}
\newcommand{\ew}{\end{widetext}}

\newcommand{\lsim}{\,\rlap{\raise 0.35ex\hbox{$<$}}{\lower 0.7ex\hbox{$\sim$}}\,}
\newcommand{\gsim}{\,\rlap{\raise 0.35ex\hbox{$>$}}{\lower 0.7ex\hbox{$\sim$}}\,}

\defcitealias{Crocker2018}{CKTC18}
\defcitealias{Crocker2020b}{Paper II}

\newcommand{\aref}[1]{\hyperref[#1]{Appendix~\ref{#1}}}

\title[Cosmic Ray Feedback I]{Cosmic rays across the star-forming galaxy sequence. I: Cosmic ray pressures and calorimetry}

\author[R. M.~Crocker et al.]{
Roland M.~Crocker,$^{1}$\thanks{E-mail: rcrocker@fastmail.fm (RMC)}
Mark R. Krumholz,$^{1,2}$
and
Todd A. Thompson$^{3}$
\\
$^{1}$Research School of Astronomy and Astrophysics, Australian National University, Canberra 2611, A.C.T., Australia\\
$^{2}$ARC Centre of Excellence for Astronomy in Three Dimensions (ASTRO-3D), Canberra 2611, A.C.T., Australia\\
$^{3}$Department of Astronomy and Center for Cosmology \& Astro-Particle Physics, The Ohio State University, Columbus, Ohio 43210, U.S.A
}

\date{Accepted XXX. Received YYY; in original form ZZZ}

\pubyear{2020}

\begin{document}
\label{firstpage}
\pagerange{\pageref{firstpage}--\pageref{lastpage}}
\maketitle

\begin{abstract}
In the Milky Way, cosmic rays (CRs) are dynamically important in the interstellar medium, contribute to hydrostatic balance, and may help regulate star formation. However, we know far less about the importance of CRs in galaxies whose gas content or star formation rate differ significantly from those of the Milky Way. Here we construct self-consistent models for hadronic CR transport, losses, and contribution to pressure balance as a function of galaxy properties, covering a broad range of parameters from dwarfs to extreme starbursts. While the CR energy density increases from $\sim 1$\,eV cm$^{-3}$ to $\sim 1$\,keV cm$^{-3}$ over the range from sub-Milky Way dwarfs to bright starbursts, strong hadronic losses render CRs increasingly unimportant dynamically as the star formation rate surface density increases. In Milky Way-like systems, CR pressure is typically comparable to turbulent gas and magnetic pressure at the galactic midplane, but the ratio of CR pressure to gas pressure drops to $\sim 10^{-3}$ in dense starbursts. Galaxies also become increasingly CR calorimetric and gamma-ray bright in this limit. The degree of calorimetry at fixed galaxy properties is sensitive to the assumed model for CR transport, and in particular to the time CRs spend interacting with neutral ISM, where they undergo strong streaming losses. We also find that in some regimes of parameter space hydrostatic equilibrium discs cannot exist, and in Paper II of this series we use this result to derive a critical surface in the plane of star formation surface density and gas surface density beyond which CRs may drive large-scale galactic winds.
\end{abstract}




\section{Introduction}

Star formation is a remarkably inefficient process: even in the cold, molecular phase of the interstellar medium, where thermal pressure support is negligible, only $\sim 1\%$ of the gas mass converts to stars per free-fall timescale \citep[e.g.,][]{Krumholz2007, Krumholz2012b, Leroy2017, Utomo2018}, or $\sim 10\%$ per galactic orbit \citep[e.g.,][]{Kennicutt1998, Kennicutt2012}. The origin of this inefficiency has long been debated, but it must at least in part be related to the various sources of non-thermal pressure that prevent the interstellar medium (ISM) from undergoing  a catastrophic free-fall collapse to the galactic midplane. The most obvious inhibitor of collapse is the supersonic turbulent motions that are ubiquitous in the interstellar media of all observed galaxies.
Turbulence
may, in turn, be driven either by mechanical feedback from supernovae, gravitational instabilities as matter flows inward through galaxies, or some combination of both \citep[e.g.,][]{Thompson2005,Ostriker2011, Faucher-Giguere2013, Krumholz2016, Hayward2017, Krumholz2018}.
Turbulence, moreover, 
naturally gives rise to a magnetic field that provides a pressure comparable to the turbulent ram pressure \citep[e.g.,][]{Federrath14a, Federrath16a}. However, in the Solar neighbourhood within the Milky Way, the midplane pressure contributed by gas motions and magnetic fields is not entirely dominant. Instead, two other sources of non-thermal pressure -- radiation and cosmic rays (CRs) -- make comparable contributions \citep{Parker1966,Boulares1990}.

While we can measure the strength of these non-thermal contributions \textit{in situ} in the Solar neighbourhood, our knowledge of their importance in galaxies with significantly-different large-scale properties (e.g., higher or lower surface densities of gas), or even elsewhere in our own Galaxy, is much more indirect and model-based. There has been significant recent theoretical progress on the importance of radiation pressure, but its role in driving turbulence and outflows in both intensely star-forming galaxies and the star clusters of normal galaxies remains uncertain \citep[e.g.,][]{Thompson2005,Andrews2011,Krumholz2012,Krumholz2013, Davis2014, Skinner15a, Tsang2015, Tsang2018, Thompson2016, Raskutti16a, Raskutti17a, Crocker2018, Crocker2018b, Wibking2018}.

The dynamical importance of CRs is even more uncertain. This is in part because most early work on this question focused only on galactic conditions similar to those found locally \citep{Jokipii1976,Badhwar1977,Ghosh1983,Chevalier1984,Boulares1990,Ko1991,Ptuskin2001}, 
and/or focused largely on the question of how and whether CRs can drive galactic winds originating in the ionised, low-density medium found several scale heights above galactic planes (\citealt{Ipavich1975,Breitschwerdt1991,Zirakashvili1996,Ptuskin1997,Zirakashvili2006}; however, for an exception see \citealt{Breitschwerdt1993}).
More recent numerical and analytic models have continued in this vein \citep[e.g.,][]{Everett2008, Jubelgas2008, Samui2010, Wadepuhl2011, Uhlig2012, Booth2013, Pakmor2016, Simpson2016, Recchia2016, Recchia2017, Ruszkowski2017, Pfrommer2017, Buck2019}, rather than address the question of whether CRs represent a significant contribution to the support of the neutral material that dominates the total mass budget and occupies at least $\sim 50\%$ of the volume \citep[e.g.][]{Dekel2019} near the midplane. Indeed, the vast majority of published simulations that include CR transport do not resolve the neutral phase or galactic scale heights ($\sim 100$ pc), and those that do \citep[e.g.,][]{Hanasz2013, Salem2014, Salem2016, Chan2019} generally assume that CR transport in the neutral ISM is identical to that in the ionised ISM \citep[though see][]{Farber2018}, an assumption that is almost certainly incorrect \citep[e.g.,][]{Zweibel2017, Xu2017, Krumholz2019}. Only a few published models attempt to address the question of CR pressure support in the neutral ISM for non-Solar neighbourhood (mostly starburst or Galactic Centre) conditions 
\citep[e.g.,][]{Thompson2006, Socrates2008, Lacki2010,Lacki2011, Crocker2011, Crocker2012, Lacki2013, Yoast-Hull2016,Yoast-Hull2019, Krumholz2019}.

Observations can provide some insight into the importance of CRs beyond the Milky Way, but thus far those efforts too have proven limited. The well-known far infrared-radio correlation \citep{Condon1992} indicates a correlation between galaxies' star formation rates and their leptonic CR populations, but since synchrotron luminosity depends not just on CR electron acceleration, but on complex factors such as the amplitude of the magnetic field and the local interstellar radiation field, it has proven challenging to draw strong conclusions about CR acceleration from radio observations alone. Several authors have argued that radio observations favour a model in which CR pressure is dynamically weak, but to date all published models have treated the interstellar medium in a simple one-zone approximation through which CR transport is described solely by parameterised timescales for escape and energy loss \citep[cf.][]{Thompson2006, Lacki2010,Lacki2013}. Moreover, radio observations directly constrain only leptonic CRs, whereas hadronic CRs (i.e., protons and heavier ions) carry the bulk of the CR energy density and pressure. Beyond the Milky Way, direct detection of $\gamma$-rays produced by the hadronic CRs that carry most of the energy has only recently become possible with the launch of the \textit{Fermi}/LAT experiment and the development of the current generation of Imaging Air Cherenkov telescope arrays \citep[e.g.,][]{Funk2015}. While there is now an established literature  -- first anticipating, more recently, contemplating \citep[e.g.,][]{Suchkov1993, Volk1996,Zirakashvili1996,Torres2004,Domingo2005,Thompson2007,Persic2008,Lacki2011,Martin2014,Yoast-Hull2016,Pfrommer2017,Sudoh2018,Peretti2019} -- the implications of the $\gamma$-ray detection of star-forming galaxies, the number of star-forming galaxies detected thus far is still $<10$ \citep[e.g.,][]{Acciari2009,Acero2009,Abdo2010,Ackermann2012,Martin2014,Rojas-Bravo2016,Fermi2019,Ajello2020,Xi2020}, and such $\gamma$-rays signals as have been detected may, in any case, be polluted by contributions from various sources or processes other than a galaxy's diffuse, hadronic CR population.\footnote{Possible contaminants include individual SNRs and/or leptonic $\gamma$-ray emission via inverse Compton or bremsstrahlung emission. Emission from AGN may also contribute in some local $\gamma$-ray detected galaxies, e.g., NGC 1068, NGC 2403, 
NGC 3424, NGC 4945, and Circinus \citep[e.g.,][]{Ajello2020}.}. 

This summary of the current state of affairs suggests that a first-principles effort to understand where and when CRs might be important, taking into account all the available observational constraints, seems warranted, and this is the primary goal of this paper. We seek to cut a broad swathe across the parameter space of star-forming galaxies, and determine where within this parameter space CRs might be dynamically significant. In a companion paper, \citepalias[Crocker et al. 2020b, hereafter][]{Crocker2020b}, we use the framework  developed here to address the closely related question: When can we expect CRs to start driving winds in the neutral interstellar media of galaxies?

The remainder of this paper is structured as follows: in \autoref{sec:setup} we present the mathematical setup of our problem and, in particular, set out the ordinary differential equation (ODE) system that describes a self-gravitating gaseous disc that maintains a quasi-hydrostatic equilibrium while subject to a flux of CRs injected at its midplane; in \autoref{sec:equilibria} we present, describe, and evaluate the numerical solutions of our ODEs; in \autoref{sec:implications} we consider the astrophysical implications of our findings for CR feedback on the dense, star-forming gas phase of spiral galaxies; we further discuss our results and summarise in \autoref{sec:discussion}. 

\section{Setup}
\label{sec:setup}

\subsection{Physical Model}

The physical system that we consider here is similar to that in \citet{Breitschwerdt1991,Breitschwerdt1993} and \citet{Socrates2008}, and which we have used in previous studies of radiation pressure feedback \citep{Krumholz2012,Krumholz2013, Crocker2018, Crocker2018b, Wibking2018}: an idealised 1D representation of a portion of a galactic disc consisting of a gas column confined by gravity through which radiation or CRs are forced from below. We are interested in exploring the equilibrium state of such a system with the goal of determining under what circumstances we expect CRs to be a significant contributor to the vertical pressure support of galactic discs.
In the companion paper \citepalias{Crocker2020b},
we determine the circumstance under which it is possible for CRs to launch winds of material out of galactic discs.
For convenience, we summarise the meanings and definitions of all symbols we introduce in this discussion in \autoref{tab:symb_defns}.

\begin{table*}
\begin{center}
    \begin{tabular}{cccc}
    \hline
    Symbol & Meaning & Defining equation & Adopted value or range\\
    \hline
    \multicolumn{4}{c}{Galactic disc parameters} \\ \hline
    $\Sigma_{\rm gas}$ & Gas surface density & & $1 - 10^{4.5}$ $M_\odot$ pc$^{-2}$ \\
    $\dot{\Sigma}_\star$ & Star formation surface density & & $10^{-4}-10^3$ $M_\odot$ pc$^{-2}$ Myr$^{-1}$ \\
    $f_{\rm gas}$ & Disc gas fraction & & $0-1$ \\
    $\sigma$ & Gas velocity dispersion & & $10-100$ km s$^{-1}$ \\
    $\beta$ & Gas velocity dispersion normalised to $c$ & $\beta=\sigma/c$ & $3\times 10^{-5} - 3\times 10^{-4}$ \\
    $\chi$ & Ion mass fraction & & $10^{-4} - 10^{-2}$ \\
    $v_A$ & Gas Alfv\'en speed & & $10-100$ km s$^{-1}$ \\
    $\mathcal{M}_A$ & Alfv\'en Mach number & $\mathcal{M}_A=\sigma/v_A$ & $1.5$ ($1-2$) \\
    $\phi_B$ & Magnetic support parameter &  \ref{eq:PhiB} & $28/27$\\
    \hline
    \multicolumn{4}{c}{CR quantities} \\ \hline
    $\kappa_{\rm conv}$ & Convective diffusion coefficient & \ref{eq:kappa_conv} \\
    $K_*$ & Midplane diffusion coefficient normalised to $\kappa_{\rm conv}$ & \ref{eq:kappa_scaling} \\
    $q$ & Index of diffusion coefficient-density relation & \ref{eq:kappa_scaling} & $1/4$ ($1/6-1/2$) \\
    $\Sigma_{\rm pp}$ & Grammage required to reduce CR flux by one $e$-folding & \ref{eq:Sigma_pp} & $1.6\times 10^5$ $M_\odot$ pc$^{-2}$ \\
    $\tau_{\rm pp}$ & Ratio of $\Sigma_{\rm gas}$ to $\Sigma_{\rm pp}$ & \ref{eq:tauppDefn} \\
    $v_s$ & CR streaming speed & \ref{eq:v_s} \\
    $\beta_s$ & CR streaming speed normalised to $c$ & $\beta_s=v_s/c$ \\
    $\epsilon_\star$ & CR energy injected per unit mass of star formation & \ref{eq:epsStar} & $5.6\times 10^{47}$ erg $M_\odot^{-1}$\\
    \hline
    \multicolumn{4}{c}{Scaling factors} \\ \hline
    $\Sigma_{\rm gas,1}$ & \multicolumn{1}{c}{$\Sigma_{\rm gas,1} = \Sigma_{\rm gas}/10$ $M_\odot$ pc$^{-2}$} & & $0.1-10^{3.5}$ \\
    $\dot{\Sigma}_{\star,2}$ & \multicolumn{1}{c}{$\dot{\Sigma}_{\star,2}=\dot{\Sigma}_\star/10^{2}$ $M_\odot$ pc$^{-2}$ Myr$^{-1}$} & & $10^2-10^5$\\
    $\sigma_1$ & \multicolumn{1}{c}{$\sigma_1=\sigma/10$ km s$^{-1}$} & & $1-10$\\
    $\chi_{-4}$ & \multicolumn{1}{c}{$\chi_{-4}=\chi/10^{-4}$} & & $1-100$ \\
    \hline \multicolumn{4}{c}{Reference (normalising) quantities} \\ \hline
    $z_*$ & Scale height & \ref{eq:z_*} \\
    $g_*$ & Gravitational acceleration & \ref{eq:g_*} \\
    $\rho_*$ & Density & \ref{eq:rho_*} \\
    $P_*$ & Pressure & \ref{eq:defnPstar} \\
    \hline \multicolumn{4}{c}{Dimensionless model quantities} \\ \hline
    $\xi$ & Height above midplane & \ref{eq:nondimensional_variables} \\
    $s$ & Column density from midplane & \ref{eq:nondimensional_variables} \\
    $p_c$ & CR pressure & \ref{eq:nondimensional_variables} \\
    $\mathcal{F}_c$ & CR flux & \ref{eq:nondimensional_variables} \\
    $r$ & Gas density & $r=ds/d\xi$ \\
    $\tau_{\rm stream}$ & Optical depth of disc to CR streaming losses & \ref{eq:tausDefn} \\
    $\tau_{\rm abs}$ & Optical depth of disc to CR absorption losses &
    \ref{eq:tausDefn} \\
    \hline
    \end{tabular}
\end{center}
    \caption{Symbol definitions
    \label{tab:symb_defns}
    }
\end{table*}

\subsubsection{Equations for transport and momentum balance}

We work in 1-dimension, $z$, the height above the midplane\footnote{By symmetry, we can just treat the half-plane from vertical height $z = 0$ to $z \to \infty$.}, and treat CRs in the fluid dynamical limit whereby they behave as a fluid of given adiabatic index $\gamma_c$; below we adopt the relativistic limit and set $\gamma_c = 4/3$. CRs are injected by supernova explosions, which we approximate as occurring solely in a thin layer near $z=0$. Adopting, e.g., Eq.~30 from \citet{Zweibel2017} \citep[also cf.][Eq.~5]{McKenzie1982,Breitschwerdt1993} to 1-dimension ($\nabla \to d/dz$) and assuming a stationary configuration ($\partial X/\partial t \to 0$ and $v_{\rm gas} \to 0$), but also now accounting for collisional energy losses of cosmic rays \citep[not included in the equation written down by][]{Zweibel2017} we have the following equation for CR transport:
\begin{eqnarray}
\label{eq:Pcr2ndOrder}
\frac{dF_c}{dz} = 
- \frac{u_c}{t_{\rm col}} + v_s \frac{d P_c}{dz} \, ,
\end{eqnarray}
in which $F_c= F_c(z)$ is the CR energy flux\footnote{Note that, in full generality, the CR energy flux contains both diffusive and advective contributions. As we explain below, however, here and, in particular, in \citet{Crocker2020b}, 
we are interested in probing
the condition of
hydrostatic equilibrium.
Thus, the systematic flow of the gas in our model is set to zero. Furthermore, while CRs may still stream with respect to the (quasi) static gas, following \citet{Krumholz2019} we shall actually treat
this motion via an effective diffusion coefficient. Altogether, these mean that below, the CR flux shall contain only a diffusive part: see \autoref{eq:Fc}.}, $u_c = u_c(z)$ is the CR energy density, $P_c = P_c(z) = (\gamma_c - 1) u_c$
is the CR pressure, 
$t_{\rm col}$ is the timescale for collisional losses, and the final term on the RHS of \autoref{eq:Pcr2ndOrder} describes exchange of energy between CRs and magnetic waves mediated by the streaming instability. Here $v_s$ is the CR streaming speed, which depends on the microphysical CR transport mechanism; we defer the question of its value for the moment, and for now simply treat it as a known quantity. We also omit second-order Fermi-acceleration, on the grounds that it is likely unimportant compared to CR escape and collisional losses \citep{Zweibel2017}.
In keeping with our assumption that all CR injection happens at $z=0$, we do not include a source term in \autoref{eq:Pcr2ndOrder}; instead, we adopt a boundary condition that $F_c$ takes on some particular non-zero value at $z=0$.

The (quasi-)hydrostatic equilibrium condition gives us a second ODE\footnote{Note that the magnetic waves launched by CR streaming provide, in principle, a yet further pressure term \citep[cf.][]{Ko1991}. 
However, given that our primary interest below is in the physical regime where ion-neutral damping quickly kills such waves, we approximate their pressure contribution as zero.
}:
\begin{equation}
\label{eq:hydroEqm}
\frac{d }{d z}\left(P_c + P_{\rm gas} + P_B - 2 P_{B_z} \right) = -\rho_{\rm gas}g_z
\end{equation}
Here $P_{\rm gas}$ is the gas pressure, $P_B = |\mathbf{B}|^2/(8 \pi)$ is the total magnetic field pressure\footnote{Note that, while it is the {\it total} magnetic field that appears, in principle, in the equation of hydrostatic balance (see \citealt{Boulares1990} and also \S 10.1.2 of \citealt{Krumholz2015}), below we shall specialise to the physically-plausible case where the turbulent magnetic field is dominant.}, $-2\, dP_{B_z}/dz = -(1/4\pi) d(|B_z|^2)/dz$ is the magnetic tension force in the vertical direction, $\rho_{\rm gas} = \rho_{\rm gas}(z)$ is the volumetric gas density, and 
\begin{equation}
g_z(z) = 4 \pi G \left[\Sigma_{\rm gas, 1/2}(z) + \Sigma_{\star, 1/2}(z)\right].
\end{equation}
is the magnitude of the acceleration in the vertical direction.
This acceleration is due to a combination of stars and gas; the gas half-column integrated from the midplane to any height $z$ is
\begin{equation}
\Sigma_{\rm gas, 1/2}(z) = \int_0^{z} \rho_{\rm gas}(z') \,dz',
\end{equation}
while the stellar half-column is $\Sigma_{\star,1/2}$. 
Consistent with our treatment of CR injection, we assume the stars are in a thin layer near $z=0$, so $\Sigma_{\star,1/2}$ is constant for all $z>0$. 
The total column of gas through the disc, i.e., including both $z<0$ and $z>0$, we denote (without the $z$ argument) as
\begin{equation}
\Sigma_{\rm gas} = \lim_{z \to \infty} 2 \Sigma_{\rm gas, 1/2}(z)
\end{equation}
and the total stellar column is $\Sigma_{\star} = 2\Sigma_{\star,1/2}$.
For future convenience, we also define the 
total gas fraction
\begin{equation}
f_{\rm gas} = \frac{\Sigma_{\rm gas}}{\Sigma_{\rm gas} + \Sigma_{\rm \star}} \, ,
\end{equation}
so the total surface mass density is 
\begin{equation}
\Sigma_{\rm tot} = \frac{\Sigma_{\rm gas}}{\fg} \, .
\end{equation}

The next step in our calculation is to adopt models for the various terms appearing in \autoref{eq:Pcr2ndOrder} and \autoref{eq:hydroEqm}; we proceed to do so in the remainder of this section.

\subsubsection{Model for gas and magnetic pressure}
\label{sec:turbulence}

Essentially all observed galaxies have neutral gas velocity dispersions that are at least transsonic (e.g., \citealt{Stilp13a, Ianjamasimanana15a, Caldu-Primo15a}; for a recent compilation, see \citealt{Krumholz2018}), so that turbulent pressure support is as or more important than thermal pressure. We must therefore adopt a model for turbulence. Given that this turbulence is injected at scales approaching the gas scale height and cascades down from there, we shall make the assumption that the turbulent velocity dispersion $\sigma$ of the gas is constant. This position-independent turbulent velocity dispersion together with the local matter density sets the  dynamical pressure within the gas column:
\begin{equation}
P_{\rm gas}(z) = \frac{2}{3} u_{\rm turb}(z) = \rho_{\rm gas}(z) \sigma^2 \, ,
\end{equation}
where $u_{\rm turb}$ is the turbulent energy density and $\sigma^2 = {\rm const}$ is the turbulent velocity dispersion. 

We further assume that the ratio of magnetic to turbulent energy is roughly constant, as expected for a magnetic field that is largely the product of a turbulent dynamo \citep[e.g.,][]{Ostriker01a, Federrath14a, Federrath16a}. Under this assumption, we can rewrite \autoref{eq:hydroEqm} as
\begin{equation}
    \frac{dP_c}{dz} + \phi_{\rm B} \sigma^2 \frac{d\rho_{\rm gas}}{dz} = -\rho_{\rm gas} g_z,
    \label{eq:hydroEqm1}
\end{equation}
where
\begin{equation}
    \phi_{\rm B} \equiv 1 + \frac{P_B - 2 P_{B_z}}{P_{\rm gas}}.
\label{eq:PhiB}
\end{equation}
The quantity $\phi_{\rm B}$ lies in the range 0 to 2, with values $>1$ indicating magnetic pressure support and values $<1$ indicating confinement by magnetic tension. 
If the turbulence is isotropic (i.e., $B_z^2 \simeq |\mathbf{B}|^2/3$), then the action of the turbulent dynamo is expected to amplify the local magnetic field amplitude such that it is close to, but perhaps slightly below, equipartition with respect to the energy density of the gas turbulent motions \citep{Federrath16a}; this implies that we have strictly $\MA \geq 1$ 
with an expected value of $\MA$ in the range $\sim 1-2$
\citep[see][]{Federrath16a,Krumholz2019}. 
Given this we have that $\phi_B$ is further restricted to the small range 1 to 13/12.
We henceforth adopt $\MA = 1.5$ and the resultant $\phi_{\rm B}  = 28/27$  as our fiducial choices for these parameters.

\subsubsection{Model for CR collisional losses}

The collisional loss time scale is 
\begin{equation}
t_{\rm col}(z) = \frac{1}{c n(z) \sigma_{\rm col} \eta_{\rm col}}
\label{eq:tcol}
\end{equation}
in which $n(z)$ is the position dependent target nucleon density, and $\sigma_{\rm col}$ and $\eta_{\rm col}$ are the total cross-section and inelasticity of the relevant collisional loss process. Given that relativistic ions dominate the energy density for reasonable assumptions about the CR distribution\footnote{Specifically, we assume that the ions follow a power law distribution in (the absolute magnitude of the) momentum \citep{Bell1978}, $p$, falling somewhat more steeply than $p^{-2}$ as a result of first-order Fermi acceleration in combination with transport timescales that also decline with momentum. CR electrons, which suffer considerably more severe losses than ions, are expected \citep[e.g.,][]{Strong2010} to constitute $\lsim$ few \% of the total cosmic ray energy density for ISM conditions in star-forming galaxies.}, and consistent with our earlier choice to set $\gamma_c \to 4/3$, we shall ignore the energetically sub-dominant, low energy, sub-relativistic cosmic ray population and treat the CRs in the relativistic limit. Given the relativistic CRs are close to or above the pion production threshold, we shall consequently assume that CR collisional losses are dominated by hadronic processes (rather than Coulomb or ionising collisions which dominate for sub-relativistic CR ions). In this case for the cross-section and elasticity $\sigma_{\rm col}$ and $\eta_{\rm col}$ in \autoref{eq:tcol} we have \citep[e.g.,][]{Kafexhiu2014}
\begin{eqnarray}
\sigma_{\rm col} \to \sigma_{\rm pp} \simeq 40 \ {\rm mbarn}  & {\rm and} & \eta_{\rm col} \to \eta_{\rm  pp} \simeq 1/2 \, .
\end{eqnarray}
Note that the hadronic collision cross-section is only weakly energy dependent above CR (proton) kinetic energies of $T_c \sim$ GeV; given that CR protons are expected to dominate the `target' and `beam' populations, we generically label these as `pp', and we set the target density to $n(z) = \rho_{\rm gas}(z)/\mu_p m_p$, where $m_p $ is the proton mass and $\mu_p \simeq 1.17$ is the ratio of protons to total nucleons for a gas that is 90\% H, 10\% He by number. For these choices, the collisional loss timescale is
\begin{equation}
    t_{\rm col} = 53 n_0^{-1}\mbox{ Myr} = 100 \rho_{{\rm gas}, -24}^{-1}\mbox{ Myr},
    \label{eq:tcol2}
\end{equation}
where $n_0 = n/1$ cm$^{-3}$ and $\rho_{{\rm gas},-24} = \rho_{\rm gas}/10^{-24}$ g cm$^{-3}$.

\subsubsection{CR fluxes}
\label{sssection:CRtrans}

The final model we must adopt is a description of how CRs interact with the magnetised turbulence in the ISM, which in turn will specify the CR flux, $F_c$. The microphysical processes responsible for scattering and confining CRs are significantly uncertain, and for this reason we will leave our analysis as generic as possible for the moment, deferring detailed models to \autoref{ssec:CR_microphysics}. We treat the flux in the standard diffusion approximation \citep{Ginzburg1964}, whereby
\begin{equation}
    F_c = -\kappa \frac{du_c}{dz}.
\label{eq:Fc}
\end{equation}
It is convenient to normalise $\kappa$ to its minimum possible value, by writing
\begin{equation}
    \kappa = K \kappa_{\rm conv},
    \label{eq:K_defn}
\end{equation}
where
\begin{eqnarray}
    \kappa_{\rm conv} & = & \frac{z_* \sigma}{3} =   \frac{\sigma^3 \ \fg}{6 \pi \ G \ \Sigma_{\rm gas}} \nonumber \\
    & \simeq & 3.8 \times 10^{26} \ \rm{cm}^2 \ \rm{s}^{-1} \ \sigma_1^3 \ \fg \ \Sigma_{\rm gas, 1}^{-1},
    \label{eq:kappa_conv}
\end{eqnarray}
$z_*$ is the gas scale height (defined precisely below), and we have defined $\sigma_1 = \sigma/10$ km s$^{-1}$ and 
$\Sigma_1=\Sigma_{\rm gas}/(10\,M_\odot\,\mathrm{pc}^2)$; the velocity dispersion and gas surface density to which we have scaled are the approximate values in the Solar neighbourhood. Here $\kappa_{\rm conv}$ is the ``convective'' diffusion coefficient that would apply if we were to assume that CRs were perfectly frozen into the gas, and were mixed solely by passive advection along with the gas, which is stirred by turbulence with a characteristic coherence length of order the galactic scale height \citep{Tennekes1972}. Since convection occurs in addition to whatever processes might allow CRs to move relative to the gas, the true diffusion coefficient is always greater than the convective one, and thus we have $K \gtrsim 1$. 

In addition to the value of $K$, we must adopt a model for its dependence on density or scale height. This is inextricably linked to the microphysical model of CR propagation that we will discuss below, but for now we note that we generically expect $K$ to rise as the density falls. This is because, as one moves out of the midplane of galaxies, magnetic fields become progressively less turbulent, more ordered, and weaker \citep{Beck2015}, presenting less of a barrier to CR propagation. Given the uncertainties of exactly how the disc-halo transition for the magnetic field occurs, we elect to follow \citet{Krumholz2019} by parameterising our ignorance: we assume that the dimensionless diffusion coefficient $K$ scales with the gas density as
\begin{equation}
    K = K_* \left(\frac{\rho_{\rm gas}}{\rho_*}\right)^{-q},
    \label{eq:kappa_scaling}
\end{equation}
where $\rho_*$ and $K_*$ are normalising factors that we are free to choose. As we discuss below in the context of our specific CR propagation models, the plausible range for the index $q$ is $q=1/6 - 1/2$. We will adopt $q=1/4$ as a fiducial choice; \citet{Krumholz2019} show that the results of CR propagation models are not highly sensitive to this choice, within the plausible physical range.

\subsection{Non-dimensionalisation}
\label{ssec:non-dimensionalisation}

We have now specified models for all terms appearing in the transport and hydrostatic balance equations. Our next step is to non-dimensionalise the equations and, in the process, extract the key dimensionless numbers that govern the system. The natural length scale for our system is the scale height of the disc imposed by turbulence,
\begin{equation}
    z_* = \frac{\sigma^2}{g_*},
\label{eq:z_*}
\end{equation}
where 
\begin{equation}
    g_* = 2 \pi G \frac{\Sigma_{\rm gas}}{\fg}
\label{eq:g_*}
\end{equation}
is the characteristic acceleration due to the matter column.\footnote{Note that, given our assumption the stars are distributed in a vanishingly thin sheet, this is the scale height of the gas distribution in the limit $\fg \to 0$. In the opposite limit, $\fg \to 1$, the scale height goes to $2 z_*$.} The length scale $z_*$ also immediately defines a characteristic density scale
\begin{equation}
    \rho_* = \frac{ \Sigma_{\rm gas}}{2 z_*} = \frac{\pi G}{\fg}  \left(\frac{\Sigma_{\rm gas}}{\sigma}\right)^2,
\label{eq:rho_*}
\end{equation}
which gives the typical density of gas near the midplane.

Other natural scales are the characteristic midplane pressure $P_*$  (with related energy density $u_* = (3/2) P_*$) is given by
\begin{equation}
P_* = g_* \rho_* z_* = \rho_* \sigma^2 = \frac{\pi G}{\fg}  \Sigma_{\rm gas}^2,
\label{eq:defnPstar}
\end{equation}
and the associated flux required if the pressure is carried by a collection of relativistic particles in the free-streaming limit 
\begin{equation}
F_* = c P_* = \frac{\pi G c}{\fg} \Sigma_{\rm gas}^2.
\label{eq:defnFstar}
\end{equation}

We now proceed to non-dimensionalise our system by defining the non-dimensional variables
\begin{eqnarray}
\xi = \frac{z}{z_*}
&
s(\xi) = \left. \frac{\Sigma_{\rm gas,1/2}(z)}{\rho_* z_*} \right|_{z= z_* \xi} \nonumber \\
p_c(\xi) = \left.  \frac{P_c(z)}{P_*} \right|_{z= z_* \xi}
&
\mathcal{F}_c(\xi) = \left.  \frac{F_c(z)}{F_*} \right|_{z= z_* \xi} \, .
\label{eq:nondimensional_variables}
\end{eqnarray}
Here $\xi$, $s$, and $p_c$ are the dimensionless height, gas (half) column, CR pressure, and flux; $ds/d\xi$ is the dimensionless gas density. The physical density is
\begin{equation}
\rho(z) = \left. \rho_* \frac{ds}{d\xi} \right|_{\xi = z/z_*}.
\end{equation}

Changing to these variables, the CR transport equation,
\autoref{eq:Pcr2ndOrder}, becomes
\begin{equation}
\label{eq:Pcr1}
\frac{d \mathcal{F}_c}{d\xi} = 
- 3 \frac{z_*}{c t_{\rm col}}p_c + 
\beta_s
\frac{dp_c}{d\xi}
\end{equation}
where $\beta_s \equiv v_s/c$. Making use of \autoref{eq:K_defn} and \autoref{eq:kappa_scaling}, the dimensionless flux is
\begin{equation}
    \mathcal{F}_c = -K_* \beta \left(\frac{ds}{d\xi}\right)^{-q} \frac{dp_c}{d\xi},
    \label{eq:dimlessFlux}
\end{equation}
where $\beta \equiv \sigma/c$. Similarly non-dimensionalising the collisional loss term (\autoref{eq:tcol}), we have
\begin{equation}
    \frac{3z_*}{c t_{\rm col}} = \frac{3\eta_{\rm pp} \sigma_{\rm pp}}{\mu_p m_p} \frac{\Sigma_{\rm gas}}{2} \frac{ds}{d\xi}.
\end{equation}
We define
\begin{equation}
\Sigma_{\rm pp} \equiv \frac{\mu_p m_p}{3 \eta_{\rm pp} \sigma_{\rm pp}} \simeq \frac{33 {\rm \ g \ cm}^{-2}}{(\eta_{\rm pp}/0.5) (\sigma_{\rm pp}/40 \ {\rm mbarn}) } \simeq 1.6 \times 10^5 \ \msun/{\rm pc}^2\,
\label{eq:Sigma_pp}
\end{equation}
as the grammage required to decrease the CR flux by one $e$-folding, so that
\begin{equation}
    \frac{3z_*}{c t_{\rm col}} = \frac{\Sigma_{\rm gas}}{2 \Sigma_{\rm pp}} \frac{ds}{d\xi} \equiv \tau_{\rm pp} \frac{ds}{d\xi}.
\label{eq:tauppDefn}
\end{equation}
Here $\tau_{\rm pp}$ is the ratio of the gas half-surface density to  $\Sigma_{\rm pp}$, which represents  the optical depth to absorption that a CR travelling in a straight line out of the galaxy would experience; we will see below that the actual optical depth to escape the galaxy is much larger than this. Inserting the quantities above into \autoref{eq:Pcr1}, and with some minor re-arrangement, we arrive at the following form of the dimensionless cosmic ray transport equation:
\begin{equation}
   \label{eq:PcrSqrdDimless}
   \frac{d}{d\xi}\left[-  \left( \frac{ds}{d\xi}\right)^{-q} \frac{dp_c}{d\xi}\right] = 
   -  \tau_{\rm abs}  \frac{ds}{d\xi} p_c  +  \tau_{\rm stream} \frac{dp_c}{d\xi},
\end{equation}
where
\begin{eqnarray}
\label{eq:tausDefn}
\tau_{\rm stream} & = & \frac{\beta_s}{K_* \beta} = \frac{1}{K_*} \frac{v_s}{\sigma} \\
\tau_{\rm abs} & = & \frac{\tau_{\rm pp}}{K_* \beta}.
\label{eq:tauaDefn}
\end{eqnarray}

Equation \ref{eq:PcrSqrdDimless} asserts that the change in CR flux with respect to height (the LHS) is equal to the rate at which CRs are lost due to collisions (the first term on the RHS) and dissipation of CR energy into Alfv\'en waves, and ultimately into thermal energy, via the streaming instability (the second term on the RHS), and we can conceptualise $\tau_{\rm abs}$ and $\tau_{\rm stream}$ as the ``absorption'' and ``streaming'' optical depths of the gas column to CRs. As noted above, the effective absorption optical depth $\tau_{\rm abs}$ is larger than the optical depth $\tau_{\rm pp}$ experienced by a CR travelling in a straight line at $c$ by a factor of $1/K_*\beta \gg 1$. This factor accounts for the fact that, although the effective speed of CRs diffusing out of the disc is $K_* \sigma$, their microphysical speed is still $c$, so the reduction in effective speed means that grammage they traverse in going a given distance must be increased by a factor $c/K_* \sigma$. 

Repeating these procedures for the equation of momentum balance, \autoref{eq:hydroEqm}, and making use of \autoref{eq:hydroEqm1}, yields the non-dimensionalised equation 
\begin{equation}
\frac{dp_c}{d\xi} + \phi_{\rm B} \frac{d^2s}{d\xi^2} = -\left(1-f_{\rm gas}\right)\frac{ds}{d\xi} - f_{\rm gas} s \frac{ds}{d\xi}.
\label{eq:HydroDimless}
\end{equation}
The terms in \autoref{eq:HydroDimless} are, from left to right, the pressure gradient due to CRs, the pressure gradient due to combined turbulence plus magnetic support, the gravitational acceleration due to stellar gravity, and the acceleration due to gas self-gravity. 

Finally, our system of \autoref{eq:PcrSqrdDimless} and \autoref{eq:HydroDimless} is fourth order in total, and thus requires four boundary conditions. 
Two of these are
\begin{eqnarray}
s(0) & = & 0 \\ \label{BC_1}
\lim_{\xi\to\infty} s(\xi) & = & 1, \label{BC_2}
\end{eqnarray}
which amount to asserting that the gas half column is zero at the midplane, and that $\lim_{z\to\infty} \Sigma_{\rm gas,1/2}(z) =  1/2 \ \Sigma_{\rm gas}$. 
For the boundary conditions on the CR pressure, 
we  can rearrange the dimensionless CR flux, \autoref{eq:dimlessFlux}, and evaluate it at $\xi =0$. This generates a third boundary condition,
\begin{equation}
\left. - 
\left(\frac{ds}{d\xi}\right)^{-q} \frac{dp_c}{d\xi} \right|_{\xi = 0} =
\frac{\tau_{\rm stream}}{\beta_s}  \frac{F_{c,0}}{F_*} \equiv f_{\rm Edd},
 \label{BC_3}
\end{equation}
where the quantity $f_{\rm Edd}$ is the ratio of the incoming CR flux to the Eddington flux,
defined here as the flux for which the momentum flux carried in the $+z$ direction by the cosmic rays matches the momentum flux in the $-z$ direction due to gravity.
Note here that
 $F_{c,0}$ is enhanced by the factor $\tau_{\rm stream}/\beta_s$ that accounts for the diffusive nature of the CR transport \citep[c.f.,][]{Socrates2008}.

To obtain the final boundary condition, 
we follow \citet{Krumholz2019} and demand that the solution of CR propagation within the disc join smoothly to the solution for free-streaming CRs as $z\to\infty$, on the basis that, once one is sufficiently high above the disc, field lines should straighten out and CRs should be able to free-stream to infinity at the Alfv\'en velocity. This condition requires that the CR enthalpy flux obey
\begin{equation}
\label{BC_4_dimfull}
 \lim_{z \to \infty} \frac{F_c}{u_c + P_c} = v_{s,\infty},
\end{equation}
where $v_{s,\infty}$ is the streaming speed well above the disc. In terms of the dimensionless parameters, this becomes
\begin{equation}
\label{BC_4}
\lim_{\xi \to \infty}  \frac{1}{\tau_{\rm stream,\infty}}   \left(\frac{ds}{d\xi}\right)^{-q} \frac{dp_c}{d\xi} = \lim_{\xi \to \infty} 4 \ {\rm sign}\left(\frac{dp_c}{d\xi}  \right) \  p_c(\xi),
\end{equation}
with $\tau_{\rm stream,\infty}$ defined identically to $\tau_{\rm stream}$, but with $v_{s,\infty}$ in place of $v_s$ (c.f.~\autoref{eq:tausDefn}). In general we expect $v_{s,\infty} > v_s$ and thus $\tau_{{\rm stream},\infty} > \tau_{\rm stream}$, because the density falls faster than the magnetic field strength as $z$ increases (though this may be compensated for by increases in the ionisation fraction with height -- see \autoref{sssec:FLRW}). However, in practice this makes little difference; numerical experimentation shows that varying the ratio $\tau_{\rm stream,\infty}/\tau_{\rm stream}$ over the range $1 - 100$ leads to $\ll 1\%$ changes in the density and pressure profiles of the resulting solutions. This is not surprising: the choice of $\tau_{\rm stream,\infty}$ sets the effective propagation speed of CRs at $z\gg 0$, but as long as this speed is large compared to the effective propagation speed near the midplane, which it is for any reasonable choice of $\tau_{\rm stream,\infty}/\tau_{\rm stream}$, the exact numerical value of $\tau_{\rm stream,\infty}$ has little effect on the results. For simplicity we will therefore adopt $\tau_{\rm stream,\infty} = \tau_{\rm stream}$ hereafter.


\subsection{CR transport models}
\label{ssec:CR_microphysics}

The values of $K$ and $v_s$ depend on the  microphysics of CR confinement, which, as noted above, are substantially uncertain. 
For this reason, we consider three possible transport models, three based on theory and one purely empirical, that differ in their predicted scalings of $\kappa$ with large-scale galaxy properties. For convenience, we collect the predicted scalings of various parameters with  galaxy properties in \autoref{tab:CR_models}, and we compare the various models in \autoref{sssec:CRtrans_comparison}.

As presaged above, for any model of CR diffusion,
convective transport sets a lower limit to the 
diffusion coefficient.
Moreover, convection is likely to be roughly the correct model 
for transport
if CRs are self-confined by the streaming instability and the medium in which they propagate is mostly ionised. 
This is because, for CRs with energies $\sim 1$ GeV, the streaming velocity is close to the Alfv\'en speed even in mostly ionised media \citep{Skilling1971, Wiener2017}.
Thus, if the turbulence is Alfv\'enic or mildly super-Alfv\'enic, per our dynamo-inspired model, convective transport will in fact dominate.

In this scenario, we trivially have $K_* = 1$ (with
the dimensional diffusion coefficient given by \autoref{eq:kappa_conv}) which implies
a maximum escape time through the gas column
\begin{eqnarray}
    t_{\rm esc,diff} & = & 100 f_{\rm gas} \frac{\sigma_1}{\Sigma_{\rm gas,1}}\mbox{ Myr}.
    \label{eq:t_esc_conv}
\end{eqnarray}
The absorption optical depth for convective
transport is
\begin{equation}
\tau_{\rm abs} = \frac{\tau_{\rm pp}}{\beta} = 1.1 \frac{\Sigma_{\rm gas,1}}{\sigma_1};
\end{equation}
this sets an upper limit to
the effective $\tau_{\rm abs}$ for the transport
modes discussed below.

\begin{table*}
    \centering
    \begin{tabular}{c@{\qquad\qquad}cccc}
        \hline
        \multirow{3}{*}{Quantity} & \multicolumn{3}{c}{CR Transport Model} \\
        & Streaming
        & Scattering
        & Constant $\kappa_*$ \\
        & (\autoref{sssec:FLRW})
        & (\autoref{sssec:scattering})
        & (\autoref{sssec:constant_k}) \\ \hline
        $K_*$ & $\frac{1}{\sqrt{2\chi} M_A^4}$ 
        & $\frac{1}{\beta} \left(\frac{G}{2f_{\rm gas}}\right)^{p/2} \left(\frac{E_{\rm CR} M_A}{e\sigma^2}\right)^p$
        & $\frac{6\pi G \Sigma_{\rm gas} \kappa_{\rm *,MW}}{f_{\rm gas} \sigma^3}$ \\[2ex]
        $v_s/\sigma$
        & $\frac{1}{\sqrt{2\chi} M_A}$
        & $\frac{1}{\sqrt{2} M_A}$
        & $\frac{1}{\sqrt{2} M_A}$ \\ [2ex]
        $\tau_{\rm stream}$ 
        & $M_A^3$
        & $\frac{\beta}{\sqrt{2} M_A} \left(\frac{G}{2 f_{\rm gas}}\right)^{-p/2} \left(\frac{E_{\rm CR} M_A}{e\sigma^2}\right)^{-p}$
        & $\frac{f_{\rm gas} \sigma^3}{6 \sqrt{2} \pi G M_A \kappa_{\rm *,MW} \Sigma_{\rm gas}}$
        \\ [2ex]
        $\tau_{\rm abs}$ 
        & $\frac{\sqrt{2\chi} M_A^4}{\beta} \tau_{\rm pp}$ 
        & $\left(\frac{G}{2 f_{\rm gas}}\right)^{-p/2} \left(\frac{E_{\rm CR}M_A}{e \sigma^2}\right)^{-p} \tau_{\rm pp}$ 
        & $\frac{f_{\rm gas} \sigma^2 c}{6\pi G \Sigma_{\rm gas} \kappa_{\rm *,MW}} \tau_{\rm pp}$ \\ \hline
    \end{tabular}
    \caption{Key dimensionless quantities for the four CR propagation models considered in this paper. In this table, $M_A$ is the Alfv\'en Mach number of the Alfv\'enic turbulent modes, $\sigma$ is the gas velocity dispersion, $\beta=\sigma/c$, $\Sigma_{\rm gas}$ is the gas surface density, $f_{\rm gas}$ is the gas fraction, $E_{\rm CR}$ is the CR energy, $p$ is the index of the turbulent magnetic field fluctuation-size relation ($1/3$ for Kolmogorov, $1/2$ for Kraichnan),  $\kappa_{\rm *,MW}$ is our fiducial Milky Way diffusion coefficient, and $\tau_{\rm pp}$ is the optical depth of the galactic disc to CRs moving in straight lines at $c$.}
    \label{tab:CR_models}
\end{table*}

\subsubsection{Streaming plus field line random walk}
\label{sssec:FLRW}

Our first model, which we will use as our fiducial choice throughout the paper, is that presented by \citet{Xu2017} and \citet{Krumholz2019}. We refer readers to those papers for full details, and here simply summarise the most important results. The motivation for this model is that the star-forming part of the interstellar medium, the part that dominates the mass budget and for which we are interested in feedback effects, is neutral rather than ionised; even by volume the neutral material occupies $\sim 50\%$ of the available space at the midplane \citep[e.g.][]{Dekel2019}, rising to near unity as one goes to more gas-rich and intensely star-forming systems. Thus, even though CRs may spend a significant portion of their lives in the ionised galactic halo (as is observed to be the case in the Milky Way), transport through the neutral ISM that dominates the mass budget is what matters for the purposes of determining whether CRs provide significant pressure support.

In a predominantly neutral medium, strong ion-neutral damping cuts off the turbulent cascade in the ISM, and decouples ions from neutrals, at scales far larger than the gyroradii of $\sim$GeV CRs. Consequently, dissipation of CR energy via streaming instability occurs into Alfv\'en waves that propagate in the ions alone, and thus have speed
\begin{equation}
    v_s = v_{A,i} = \frac{v_A}{\sqrt{\chi}} = \frac{\sigma}{\sqrt{2 \chi} M_A}
\label{eq:v_s}
\end{equation}
where $\chi$ is the ionisation fraction by mass, $M_A$ is the Alfv\'en Mach number of the turbulence in the ISM, and the factor two in the denominator of the last term arises from the assumption that Alfv\'enic modes carry half the turbulent energy. As noted above, dynamo models predict $M_A\simeq 1-2$. We adopt $M_A=1.5$ as a fiducial choice unless noted otherwise, but explore this dependence below.

Since the external turbulence does not couple to CRs, CR transport in such a medium occurs predominantly by CRs streaming along field lines at the ion Alfv\'en speed\footnote{For CR energies $\gg$ GeV, given a reasonable power-law spectral distribution, the energy density of CRs available to excite magnetic field fluctuations at a given gyroradius scale declines sufficiently that the balance between streaming instability and ion-neutral damping no longer implies a streaming speed that is very close to the Alf\'en speed. At this point the streaming velocity then starts to grow again with energy; see \citet{Krumholz2019}. However, in this paper we focus solely on the $\sim$GeV CRs that dominate the CR pressure, and these are essentially always in the regime where the streaming speed is close to the Alfv\'en speed: see Appendix  \ref{app:streamSpeed}. Also note that it is an assumption of the streaming model that the CRs' motion  along the field lines will act to transport them down their gradient such that they can excite the streaming instability; given the overall anisotropy of our setup (with the CR sources concentrated in the plane), it is clear that, globally, CRs must move down their gradient by escaping out of the disk. However there may be local instances where the gradient criterion is not satisfied forming ``bottleneck" regions. This effect has been investigated numerically by \citet{Wiener2013,Wiener2017}.}, coupled with the random walk of those field lines in the overall turbulence, implying an effective mean free path equal to the magnetic field coherence length. The corresponding diffusion coefficient is therefore
\begin{equation}
    \kappa = \frac{v_{A,i} l_{\rm coh, B}}{3},
\end{equation}
where $l_{\rm coh, B}$ is the coherence length of the magnetic field, which for a dynamo-generated field is
\begin{equation}
l_{\rm coh, B} \simeq \frac{z_*}{M_A^3}.
\label{eq:lcoh}
\end{equation}

Consequently,  for this model we adopt
\begin{equation}
    K_* = \frac{1}{\sqrt{2 \chi} M_A^4} \simeq \frac{22.4}{\chi_{-3}^{1/2} M_A^{4}}
    \label{eq:K_FLRW}
\end{equation}
It immediately follows that\footnote{Note that, for this streaming case,
that the optical depth to {\it scattering}
is given by $z_*/l_{\rm coh,B} = M_A^3$ which is identically equal to $\tau_{\rm stream}$.
Thus for $M_A \geq 1$ we are safely in the diffusive regime.
}
\begin{eqnarray}
    \label{eq:tau_s_stream}
    \tau_{\rm stream} & = & M_A^3 \\
    \label{eq:tau_a_stream}
    \tau_{\rm abs} & = & \frac{\sqrt{2\chi} M_A^4}{\beta} \tau_{\rm pp} = 0.043 \frac{M_A^4 \Sigma_{\rm gas,1}}{\chi_{-3}^{1/2} \sigma_1},
\end{eqnarray}
where $\chi_{-3} = \chi/10^{-3}$; the ionisation fraction to which we have chosen to normalise is intermediate between the values of $\sim 10^{-4}$ found in starbursts \citep{Krumholz2019} and the value $\sim 10^{-2}$ found in the warm atomic medium of galaxies like the Milky Way \citep{Wolfire2003}. 
Thus $\tau_{\rm abs}$ is somewhat less than unity for Milky Way-like parameters, but becomes larger than unity for galaxies with larger gas surface densities. As discussed in \citet{Krumholz2019}, the value of $q$ for this model, which specifies the density scaling, is uncertain because it depends on how the ionisation fraction and coherence length of the magnetic field vary with height. However, \citet{Krumholz2019} also show that their results are not terribly sensitive to this choice.

For reference, the corresponding dimensional diffusion coefficient is\footnote{Note that
\citet{Farber2018} present numerical MHD
simulations where they try to incorporate the effect of ion-neutral damping on CR transport in neutral ISM gas
via the expedient of a diffusion coefficient that {\it increases} by a factor of ten 
to $3.0 \times 10^{28}$ cm$^2$ s$^{-1}$
in gas below a temperature of $10^4$ K.
However, we find that transport is not necessarily faster for the ``Streaming" case than for the ``Scattering" case; in general this depends on $\chi$, $\fg$, and other properties, as can be seen by comparing \autoref{eq:kappaFLRW} to the equivalent expression for scattering derived below, \autoref{eq:kappaScat}. Over the range of properties explored by observed galaxies, one can find regimes where both scattering and streaming give larger diffusion coefficients. We also find that, for the range of parameters we expect to encounter in galaxies, the ``Streaming" diffusion coefficient is substantially lower than \citeauthor{Farber2018}'s assumed $3.0 \times 10^{28}$ cm$^2$ s$^{-1}$. 
}
\begin{equation}
    \kappa_* \simeq 8.5 \times 10^{27} \frac{\sigma_1^3 \fg M_A^{-4} }{\sqrt{\chi_{-3}} \ \Sigma_{\rm gas,1}} \ {\rm cm}^2 \ {\rm s}^{-1},
    \label{eq:kappaFLRW}
\end{equation}
and the diffusive escape time is
\begin{equation}
    t_{\rm esc,diff} = \frac{z_*^2}{2\kappa_*} \simeq 4.9 \frac{M_A^4 \fg \chi_{-3}^{1/2}  \sigma_1}{\Sigma_{\rm gas,1}}  \ {\rm Myr}.
    \label{eq:tdiff}
\end{equation}
For comparison, note that the collisional loss timescale (\autoref{eq:tcol}) is
\begin{equation}
    t_{\rm col} = \frac{\mu_p m_p}{c \rho_* \sigma_{\rm col} \eta_{\rm col}} = 110\, f_{\rm gas} \left(\frac{\sigma_1}{\Sigma_{\rm gas,1}}\right)^2\,\mathrm{Myr}.
    \label{eq:t_col}
\end{equation}
Thus for Milky Way-like parameters the collisional loss time is substantially longer than the diffusive escape time, and most CRs do not produce observable $\gamma$-ray emission. However, given the generic dependencies $t_{\rm col} \propto \Sigma_{\rm gas}^{-2}$ and $t_{\rm esc,diff} \propto \Sigma_{\rm gas}^{-1}$, collisional losses will always win out over diffusive escape at sufficiently high gas surface density (for other parameters held fixed). This same point will apply equally to all the models we consider.

It is also interesting to compare these two timescales to the timescale for loss of CR energy due to damping via streaming instability. By analogy with $t_{\rm col}$ in \autoref{eq:Pcr1}, we can define the characteristic streaming loss time as
\begin{equation}
    t_{\rm stream} = \frac{3 z_*}{v_s} = 4.9  \frac{ M_A f_{\rm gas} \chi_{-3}^{1/2}\sigma_1}{\Sigma_{\rm gas,1}} \mbox{ Myr}
\end{equation}
Thus we see that, for the fiducial parameters for this model, in a Milky Way-like galaxy the streaming loss timescale is comparable to the escape time and much smaller than the collisional loss time. However, this conclusion is very sensitive to the assumed Alfv\'en Mach number ($t_{\rm esc,diff}/t_{\rm stream} \propto M_A^{3}$). Moreover, the streaming loss timescale has the same dependence on $\Sigma_{\rm gas}$ as the escape timescale, and so collisional losses increase in importance relative to streaming losses as one moves to higher surface density galaxies.

\subsubsection{Scattering off extrinsic turbulence}
\label{sssec:scattering}

Our second theoretically-motivated model is intended to apply in ionised regions. Roughly half the volume at the midplane of discs of normal galaxies is ionised, and this fraction rises as one moves into the halo, where CRs spend much of their time. Thus, despite the fact that we are mainly interested in the feedback effects of CRs in the neutral ISM, we must consider the possibility that CR propagation is mainly through the ionised phase, and that the force applied by CRs to the neutral ISM occurs primarily at the neutral-ionised interface. In an ionised gas, the turbulent cascade in the magnetic field does reach down to the CR gyroradius; however, there is a great deal of uncertainty about whether CRs are confined primarily by Alfv\'en waves that they themselves create via the streaming instability, or primarily by waves cascading from larger scales, or some combination of both \citep[e.g.,][and references therein]{Zweibel2017,Blasi2019}. If CRs are predominantly self-confined, then the transport mechanism is much the same as for the case of predominantly neutral medium, simply with the ionisation fraction $\chi = 1$. On the other hand, if they are confined by scattering off the ambient turbulence in the ISM, then we can compute the resulting diffusion coefficient for highly relativistic CRs, following, e.g., \citet{Jokipii1971} or \citet{Lacki2013}, as
\begin{equation}
    \kappa \simeq \frac{c r_g^p z_*^{1-p}}{3},
\end{equation}
where we have assumed that $z_*$ is the outer scale of the turbulence,
\begin{equation}
    r_g \simeq \frac{E_{\rm CR}}{2 e B},
\end{equation}
is the CR gyroradius (assuming a mean sin pitch angle of $1/2$), and $p$ depends on the index of the turbulent spectrum: $p=1/3$ for a Kolmogorov spectrum and $p=1/2$ for a Kraichnan spectrum. We will adopt $p=1/2$ as a fiducial choice, and use this value for all numerical evaluations; however, we give results for general $p$. The factor $q$ that describes the density scaling is $q = p/2$, i.e., for our fiducial $p=1/2$, we have $q=1/4$, so the diffusion coefficient decreases with density as $\kappa \propto \rho^{-1/4}$.

We are interested in evaluating this near the midplane, where the characteristic magnetic field strength in the ISM, $B_*$, is given by
\begin{equation}
    B_* = \frac{\sqrt{2\pi\rho_*}}{M_A} \sigma = \sqrt{\frac{2 \ G}{\fg}}\frac{ \pi \ \Sigma_{\rm gas}}{M_A}
    \label{eq:Bstar}
\end{equation}
Making this substitution, with a bit of algebra we obtain
\begin{equation}
    K_* \simeq \frac{1}{3 \beta} 
    \left(\frac{E_{\rm CR} M_A}{e \sigma^2} \sqrt{\frac{G}{2 f_{\rm gas}}}\right)^p
    \simeq 
    0.25 \, M_A^{1/2} E_{\rm CR,0}^{1/2} f_{\rm gas}^{-1/4} \sigma_1^{2}
\end{equation}
where $E_{\rm CR,0} = E_{\rm CR}/1$ GeV (and we have adopted the fiducial $p = 1/2$ for the numerical evaluation). However, note the general restriction that $K_* \geq 1$, since this is the limit set by convective transport of CRs; thus galaxies with low velocity dispersions will be in this convective limit.

As discussed by \citet{Zweibel2017} among others,  it is important to distinguish between the case where the turbulent Alfv\'en waves that scatter CRs is balanced, i.e., roughly equal power in Alfv\'en waves propagating in both directions along a field line, and unbalanced, where the Alfv\'en waves are predominantly in a single direction. In the latter case the CRs can stream with the Alfv\'en waves (although the transport is still dominated by scattering rather than streaming, i.e., streaming does little to increase the value of $K_*$), and streaming losses occur. In the former, streaming losses due to Alfv\'en waves propagating in one direction are compensated by energy gain from waves propagating in the opposite direction, and there is no net streaming loss; indeed, there may be a net gain of energy by the CRs due to second-order Fermi acceleration. For the present CR transport model, we are interested in a case where the majority of the stellar feedback driving the turbulence is injected near the midplane. Thus we will assume that the Alfv\'en modes in the turbulence are unbalanced, with waves leaving the midplane predominating. In this case the effective speed that determines the streaming loss is $v_s = v_A$, and the streaming\footnote{Note that, for this scattering case -- and for the case of constant diffusion coefficient outlined below --
the optical depth to {\it scattering}
is given by $z_*/\lambda_{\rm mfp} = 
c z_*/(3 \kappa_*) = 1/\beta_A \tau_{\rm stream}$.
Thus, while we shall find below that $\tau_{\rm stream} \lsim 1$ for parameters apposite to real galaxies and for the ``scattering" and constant $\kappa_*$ cases, at the same time, we find scattering optical depths $\gsim 1000$ and $\gsim 100$
for these two cases, respectively so we are, again, well into the diffusive regime.
The CR optical depth to scattering is
the direct analogue to
what \citet{Socrates2008} label $\tau_{\rm CR}$.
}
and absorption optical depths are therefore
\begin{eqnarray}
    \label{eq:tau_s_scat}
    \tau_{\rm stream}  &=& \frac{\beta}{\sqrt{2} M_A} 
    \left(\frac{E_{\rm CR} M_A}{e \sigma^2} \sqrt{\frac{G}{2 f_{\rm gas}}}\right)^{-p}\nonumber \\
     &\simeq& 0.96 \, M_A^{-3/2} E_{\rm CR,0}^{-1/2} f_{\rm gas}^{1/4} \sigma_1^2
\end{eqnarray}
and
\begin{eqnarray}
    \label{eq:tau_a_scat}
    \tau_{\rm abs} &=& \tau_{\rm pp} \left(\frac{E_{\rm CR} \ M_A}{e \sigma^2} \sqrt{\frac{G}{2 f_{\rm gas}}}\right)^{-p}\nonumber \\
     &\simeq& 1.3 \, M_A^{-1/2} E_{\rm CR,0}^{-1/2} f_{\rm gas}^{1/4} \sigma_1 \Sigma_{\rm gas,1}.
\end{eqnarray}
Again, note that these expressions are valid for $K_* > 1$.

For this model the dimensional diffusion coefficient and escape time, for $K_* > 1$, are
\begin{eqnarray}
\label{eq:kappaScat}
\kappa_* & = & 9.4 \times 10^{25} \, M_A^{1/2} E_{\rm CR,0}^{1/2} f_{\rm gas}^{3/4} \frac{\sigma_1}{\Sigma_{\rm gas,1}} \, \mathrm{cm}^2\,\mathrm{s}^{-1}  \\
 t_{\rm esc,diff} & = & 440 \, M_A^{-1/2} E_{\rm CR,0}^{-1/2} f_{\rm gas}^{5/4}\frac{\sigma_1^3}{\Sigma_{\rm gas,1}} \,\mathrm{Myr}.
\end{eqnarray}
The corresponding values for $K_* = 1$ are given by \autoref{eq:kappa_conv} and \autoref{eq:t_esc_conv}, respectively. The streaming loss time is
\begin{equation}
    t_{\rm stream} = \frac{3z_*}{v_{A}} = 150 \frac{M_A f_{\rm gas}  \sigma_1}{\Sigma_{\rm gas,1}}\mbox{ Myr},
\label{eq:tstreamScatt}
\end{equation}
and the collisional loss time is independent of the CR transport model (\autoref{eq:t_col}). Thus in this CR transport model streaming losses occur a factor of a few more slowly than collisional losses even for Milky Way-like conditions, and become even less important in higher surface density galaxies.

\subsubsection{Constant diffusion coefficient}
\label{sssec:constant_k}

Our final, purely empirical, model is simply to plead ignorance as to the true value of the diffusion coefficient as a function of galaxy properties, and adopt the empirically-determined Milky Way one for all galaxies: $\kappa_* \approx  \kappa_{\rm *,MW} \equiv 1\times 10^{28}$ cm$^2$ s$^{-1}$, as estimated empirically for $\sim$GeV CRs in the Milky Way \citep[e.g.,][]{Ptuskin2006}. In our dimensionless variables, this corresponds to
\begin{equation}
   K_* = \frac{\kappa_{\rm *,MW}}{\kappa_{\rm conv}} \simeq 2.6 \frac{\Sigma_{\rm gas,1}}{f_{\rm gas} \sigma_1^3}.
\end{equation}
Note that this assumption can produce $K_* < 1$, which is unphysical, but we do not enforce this condition for the purposes of comparing to previous works in which $\kappa_*$ has been treated as constant. For this model we also adopt $v_s = v_A$, in which case we have
\begin{eqnarray}
\tau_{\rm stream} & = & \frac{1}{\sqrt{2} M_A K_*} = 0.27 \frac{f_{\rm gas} \sigma_1^3}{M_A \Sigma_{\rm gas,1}} \\
\tau_{\rm abs} & = & \frac{\tau_{\rm pp}}{K_* \beta} = 0.37 f_{\rm gas} \sigma_1^2.
\end{eqnarray}
The diffusive escape time is
\begin{equation}
    t_{\rm esc,diff} = 21 f_{\rm gas}^2 \frac{\sigma_1^4}{\Sigma_{\rm gas,1}^2}\mbox{ Myr},
\end{equation}
and the streaming timescale is identical to that in the scattering model (\autoref{eq:tstreamScatt}).

\subsubsection{Comparison of transport models}
\label{sssec:CRtrans_comparison}

Before proceeding to apply the various CR transport models, it is helpful to develop some intuition by comparing their predictions for the key dimensionless ($K_*$, $\tau_{\rm abs}$, $\tau_{\rm stream}$) and dimensional ($\kappa_*$, $t_{\rm esc,diff}$, $t_{\rm col}$, $t_{\rm stream}$) parameters that describe the system as a function of galaxy gas surface density. Since these quantities also depend on additional quantities such as the gas velocity dispersion and gas fraction, it is helpful to consider a few cases that are representative of different types of galaxies. We consider three parameter sets, which we can imagine as describing typical values in local spiral galaxies, starburst / merger systems, and a case intermediate between these extremes. We summarise the parameters we adopt for these three cases in \autoref{tab:galparams}. In all cases we adopt $M_A = 1.5$ and $E_{\rm CR} = 1$ GeV.

\begin{table}
    \centering
    \begin{tabular}{c@{\qquad}ccc}
        \hline
        \multirow{2}{*}{Quantity} & \multicolumn{3}{c}{Galaxy Model} \\
        & Local & Intermediate & Starburst \\ \hline
        $\Sigma_{\rm gas}$ [$M_\odot$ pc$^{-2}$] &
        $10^0-10^{2.5}$ & $10^1-10^{3.5}$ & $10^{2.5}-10^{4.5}$ \\
        $\sigma$ [km s$^{-1}$]
        & 10 & 30 & 100 \\
        $f_{\rm gas}$ & 0.1 & 0.4 & 0.7 \\
        $\chi$ & $10^{-2}$ &
        $10^{-3}$ & $10^{-4}$ \\
        \hline
    \end{tabular}
    \caption{Example galaxy parameters. The range given for $\Sigma_{\rm gas}$ is the approximate range in galaxy surface densities over which the indicated parameter sets are plausible.}
    \label{tab:galparams}
\end{table}

\begin{figure*}
    \centering
    \includegraphics[width=\textwidth]{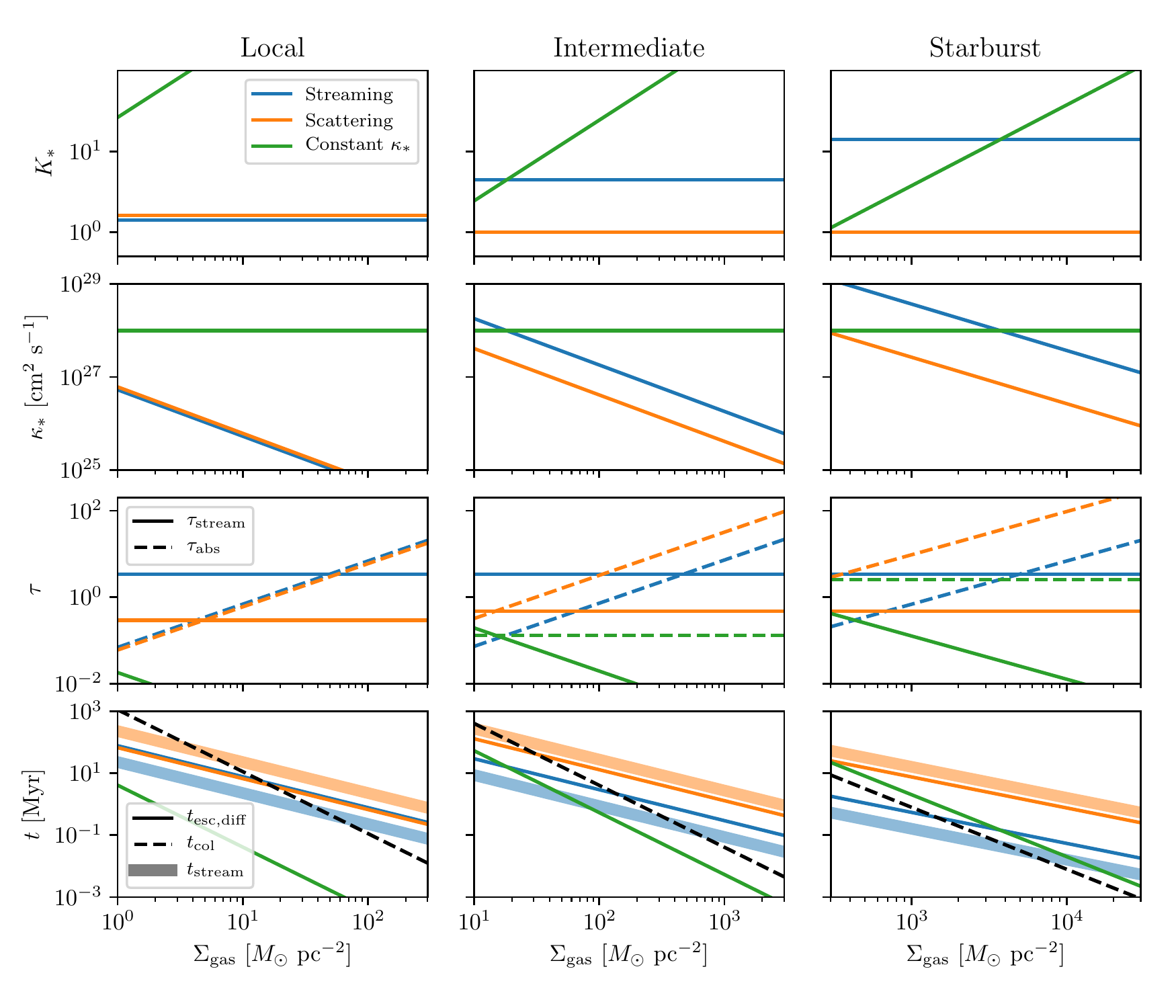}
    \caption{Dimensionless ($K_*$, $\tau_{\rm stream}$, $\tau_{\rm abs}$) and dimensional ($\kappa_*$, $t_{\rm esc,diff}$, $t_{\rm col}$, $t_{\rm stream}$) quantities as a function of gas surface density predicted by our CR transport models. The three columns are for the local, intermediate, and starburst cases, whose parameters are given in \autoref{tab:galparams}. Note that the horizontal axis range is different for each column; we have limited to axis range to gas surface densities that are reasonably plausible for each particular set of parameters.}
    \label{fig:cr_transport_models}
\end{figure*}

We plot dimensionless and dimensional parameters for our CR transport models in \autoref{fig:cr_transport_models}. The figure allows a few immediate observations. First focus on the top two rows, showing $K_*$ and $\kappa_*$. The streaming and scattering models give nearly identical values of $K_*$ and $\kappa_*$ for local galaxy conditions. However, the two models change in different directions as we shift from the local to the starburst regime: a scattering model predicts less and less efficient diffusion in higher surface density galaxies, eventually saturating at the convection limit, while the streaming model predicts more rapid transport in starburst galaxies due to the higher neutral fraction, and thus higher streaming speed, in these galaxies interstellar media. The constant $\kappa_*$ model is qualitatively different. For the other models, as the gas surface density rises, reducing the scale height and increasing the density, the CR diffusion coefficient goes down. If one assumes constant $\kappa_*$, this does not happen, and $K_*$ can be far larger or smaller than the convective value. The former is certainly unphysical, and the latter is likely unrealistic as well, and thus we will not consider the constant $\kappa_*$ model further in this work.

Now consider the lower two rows, which show the dimensionless scattering and absorption optical depths, and the loss times. Again, we can make a few immediate observations. At higher gas surface densities, $\tau_{\rm abs}$ always becomes larger than unity, and $t_{\rm col}$ smaller than $t_{\rm esc,diff}$ or $t_{\rm stream}$. Thus we expect galaxies to become increasingly calorimetric and dominated by collisional losses as we go from low to high gas surface density, with a transition to calorimetry occurring at $\sim 100-1000$ $M_\odot$ pc$^{-2}$ depending on model parameters. The sole exception to this is if one assumes constant $\kappa_*$, in which case the ratio of escape to collisional loss time is independent of gas surface density; again, this is unphysical. A second result of note is the relative size of the streaming optical depth $\tau_{\rm stream}$ and streaming loss time $t_{\rm stream}$ in the different models. Streaming losses are strongest in the streaming case, and thus should play a significant role over nearly all of parameter space. They are comparably much less important for a scattering transport model.

\section{Cosmic Ray Equilibria}
\label{sec:equilibria}

Having obtained our dimensionless equations and considered the microphysics of CR transport, we now proceed to explore the properties of CR equilibria. 

\subsection{Numerical method}

While it is possible to solve \autoref{eq:PcrSqrdDimless} and \autoref{eq:HydroDimless} analytically in certain limiting cases, in general they must be solved numerically. Our first step is therefore to develop an algorithm to obtain solutions. Because the boundary conditions, \autoref{BC_1} - \autoref{BC_4}, are specified at different points (two at $\xi=0$ and two at $\xi=\infty$), the system is a boundary value problem, which we must solve iteratively. 

Our first step is to make a change of variables to a form that renders the system somewhat more stable for numerical integration. We use as our integration variables $s$, $\ln (ds/d\xi) \equiv \ln r$, $\ln p_c$, and $f_c \equiv \mathcal{F}_c / K_* \beta p_c$; intuitively, these quantities are the column density, the logarithmic volume density, the logarithmic CR pressure, and the effective CR propagation speed. In terms of these variables, \autoref{eq:PcrSqrdDimless}, \autoref{eq:HydroDimless}, and their boundary conditions (\autoref{BC_1} - \autoref{BC_4}) become
\begin{equation}
    \frac{d}{d\xi}\left(
    \begin{array}{c}
    s \\ \ln r \\ \ln p_c \\ f_c
    \end{array}\right)
    =
    \left(
    \begin{array}{c}
    r \\
    \phi_{\rm B}^{-1}\left(f_{\rm gas} - 1 - f_{\rm gas} s + p_c r^{q-1} f_c\right) \\
    -r^q f_c \\
    -\tau_{\rm abs} r - \tau_{\rm stream} r^q f_c + r^q f_c^2
    \end{array}
    \right),
    \label{eq:numerical_eqns}
\end{equation}
with boundary conditions $s(0) = 0$, $\lim_{\xi\to\infty} s(\xi) = 1$, $f_c(0) = f_{\rm Edd} / p_c(0)$, and $\lim_{\xi\to\infty} f_c(\xi) = 4\tau_{\rm stream,\infty}$.

We then solve this system using a shooting algorithm: we have $s(0)=0$ from \autoref{BC_1}, and we start with an initial guess for the mid-plane log density $\ln r(0)$ and CR pressure $\ln p_c(0)$. These choices together with \autoref{BC_3} allow us to compute the effective propagation speed $f_c(0)$ at the midplane, so that we now have a set of four initial values at $\xi=0$. We can then integrate the system outward toward $\xi\to \infty$, stopping when either (1) $s$ and $f_c$ both approach constant values, or (2) $\ln r$ or $\ln p_c$ diverge to negative infinity, or (3) $f_c$ diverges to positive infinity. The integration must be carried out with care, since at large $\xi$ the system becomes extremely sensitive to numerical noise; we use a fourth-order implicit Runge-Kutta method to maintain stability. We then carry out a double-iteration procedure: we hold $\ln r(0)$ fixed and iteratively adjust $\ln p_c(0)$ until we find a value such that $\lim_{\xi\to\infty} f_c = 4\tau_{\rm stream}$ (\autoref{BC_4}). This choice will not in general satisfy the condition that $\lim_{\xi\to\infty} s(\xi) = 1$ (\autoref{BC_2}), and thus we next iteratively adjust $\ln r(0)$ until this boundary condition is satisfied. We note that, for sufficiently large $f_{\rm Edd}$, the procedure does not converge, and it is not possible to find a solution that satisfies the boundary conditions. We defer further discussion of this case to the companion paper, CKT20b.

\subsection{Gas density and cosmic ray pressure profiles}
\label{subsec:profiles}

\begin{figure}
\includegraphics[width=\columnwidth]{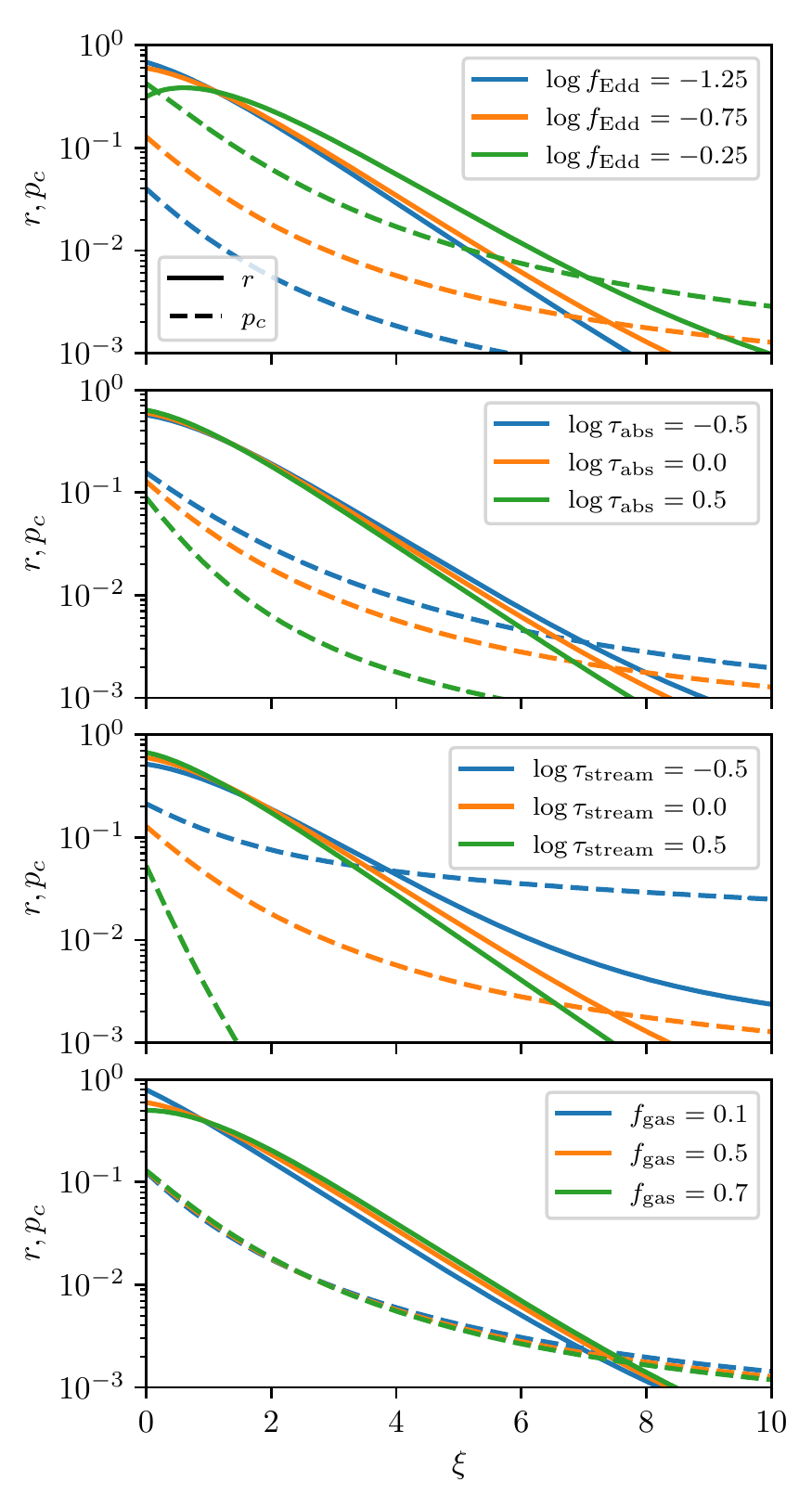}
\caption{Profiles of (dimensionless) volumetric density $r(\xi)$ (solid) and (dimensionless) CR pressure $p_c(\xi)$ (dashed). In each panel, one of the four fundamental parameters -- $f_{\rm Edd}$, $\tau_{\rm abs}$, $\tau_{\rm stream}$, and $f_{\rm gas}$ -- is varied (as indicated in the legend), while the other three are held constant; the constant values we adopt are $f_{\rm Edd} = 10^{-0.25}$, $\tau_{\rm abs} = 1$, $\tau_{\rm stream} = 1$, and $f_{\rm gas} = 0.5$.
\label{plotDensity}
}
\end{figure}

Our next step, now that we have an algorithm to generate solutions, is to develop some intuition for the behaviour of solutions and their dependence on the four fundamental parameters for our system: $\tau_{\rm stream}$ (\autoref{eq:tausDefn}), $\tau_{\rm abs}$ (\autoref{eq:tauaDefn}), $f_{\rm Edd}$ (\autoref{BC_3}), and $f_{\rm gas}$. We plot example dimensionless gas density and CR pressure profiles in \autoref{plotDensity}. In each of the four panels shown, we vary one quantity, as indicated in the legend, while holding the other three constant; the quantities not indicated in the legend have values $\log f_{\rm Edd} = 10^{-0.75}$, $\tau_{\rm abs} = 1$, $\tau_{\rm stream} = 1$, and $f_{\rm gas} = 0.5$, and in all cases we adopt our fiducial values $\phi_{\rm B} = 1.01$ and $q=1/4$. The range of parameters we have chosen are representative of the range found in observed galaxies, as we discuss below.

We can understand the results shown in each of the panels intuitively. In the top panel, we see that smaller values of $f_{\rm Edd}$ yield (not surprisingly) smaller CR pressures, and density profiles that are close to the values that would be obtained absent CR pressure. As $f_{\rm Edd}$ rises, the density profile becomes more extended and develops a long tail at high $\xi$ that is supported by CR pressure \citep[cf.][]{Ghosh1983,Chevalier1984,Ko1991}. At the highest $f_{\rm Edd}$, a mild density inversion appears near $\xi = 0$. We show in \aref{app:parker} that in such regions the solution becomes Parker unstable, but that this is unlikely to significantly modify any of our conclusions. We therefore ignore Parker stability considerations for the remainder of the main text.

Turning to the second and third panels, we see that $\tau_{\rm abs}$ and $\tau_{\rm stream}$ mainly control how rapidly the CR pressure drops with $\xi$ -- larger opacities lead to sharper drops, as more and more CRs are lost to absorption or streaming. The value of $\tau_{\rm stream}$ has more dramatic effects than the value of $\tau_{\rm abs}$, because $\tau_{\rm stream}$ not only controls streaming losses, it controls the boundary condition at infinity:  smaller $\tau_{\rm stream}$ corresponds to a smaller ratio of $F_c/(u_c + P_c)$ (i.e., less flux per unit CR enthalpy) as $z\to\infty$. Thus smaller $\tau_s$ implies larger CR pressure and energy at large $z$.

Finally, we see that the gas fraction has relatively small effects on either the density or CR pressure profiles. More gas-rich systems experience more flattening of the density profile near $\xi = 0$ as a result of CR pressure support; this is simply a consequence of the fact that the gravitational acceleration builds up more slowly with $\xi$ for larger $f_{\rm gas}$, and thus gravity is weaker near the midplane, making it easier for CR pressure to flatten the density profile.

\subsection{Cosmic ray pressure contribution and calorimetry}
\label{ssec:crpressure_example}

\begin{figure}
\includegraphics[width=\columnwidth]{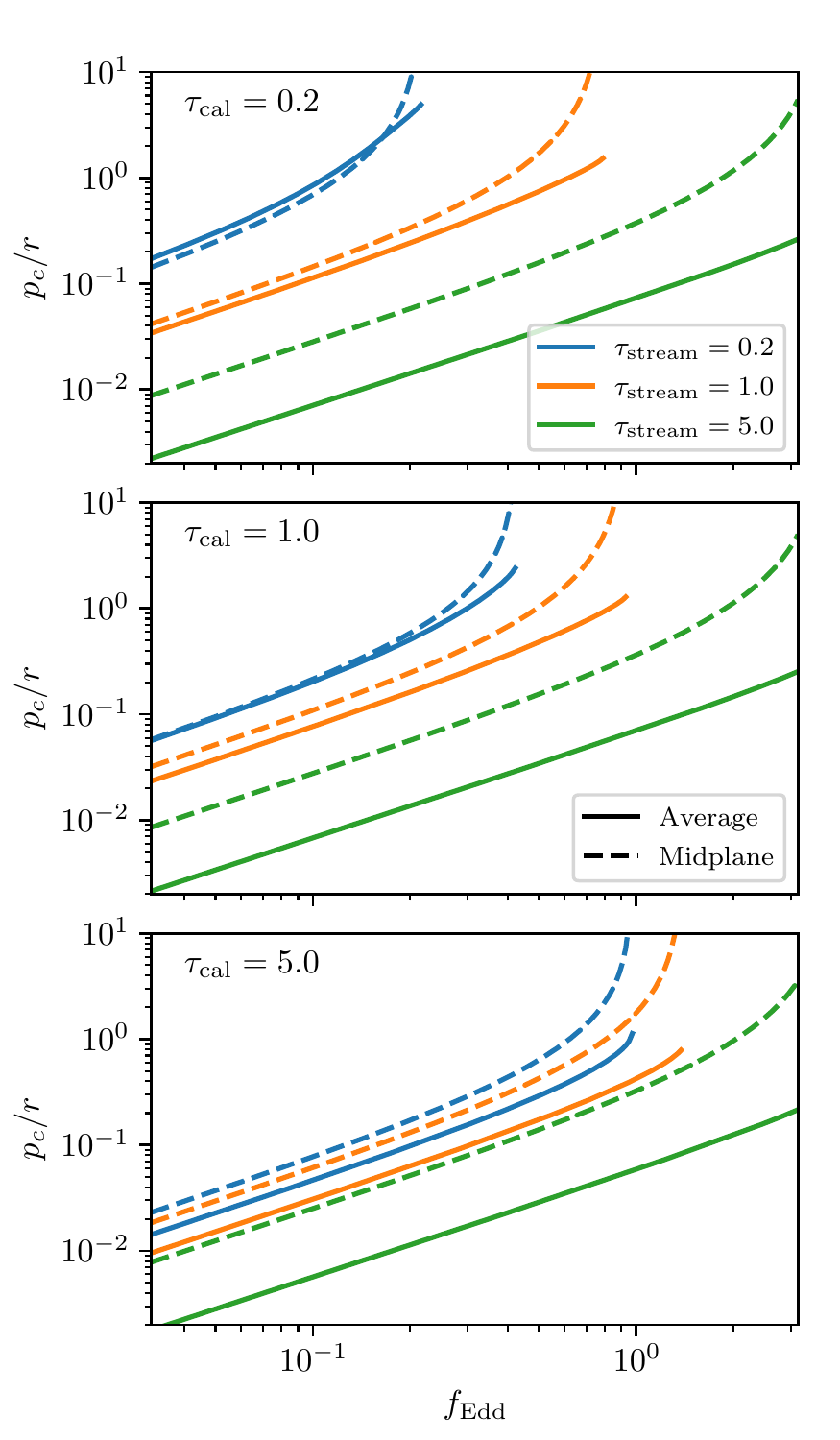}
\caption{Ratio of CR pressure to gas turbulent pressure $p_c/r = P_c/P_{\rm gas}$ as a function of $f_{\rm Edd}$, for absorption optical depths $\tau_{\rm abs}=0.2$, $1.0$, and $5.0$ (top to bottom panels), and for $\tau_{\rm stream}=0.2$, $1.0$, and $5.0$ (colors, as indicated in the legend). All cases shown use $f_{\rm gas}=0.5$, but results are qualitatively similar for any $f_{\rm gas}$. In each case, dashed lines show the ratio measured at the midplane, while solid lines show the ratio averaged over the first gas scale height.
\label{plotPressureRatioForCRpaper}
}
\end{figure}

In this paper we are interested in where CRs are dynamically important for helping support the interstellar medium, and we are now in a position to answer this question in the context of our models. \autoref{plotPressureRatioForCRpaper}
shows the ratio of CR to turbulent pressure computed for a sample of parameters. We show this ratio computed in two ways: the midplane value (dashed lines), and the average value of the first gas scale height (solid lines). For this purpose we define a scale height to be the value of $\xi$ for which $s(\xi) = 1 - e^{-1}$, i.e., the scale height is defined as the height for which the fraction of the gas mass below that height has the same value as it would at one scale height in an exponential atmosphere. Clearly we see that, as $f_{\rm Edd}$ is dialled upwards, the CRs make a larger and larger contribution to the total pressure, becoming dominant at sufficiently large $f_{\rm Edd}$; indeed, for sufficiently large $f_{\rm Edd}$, the midplane density drops to zero (as indicated by the dashed lines in the figure diverging to infinity), and no hydrostatic equilibrium is possible, a topic to which we return in \citetalias{Crocker2020b}. We truncate the lines in the plot when this condition occurs. We also see that the CR pressure contribution drops as we increase the optical depth due to the increasing importance of losses, particularly as one moves away from the midplane. For the highest optical depth cases shown in \autoref{plotPressureRatioForCRpaper}, the ratio of CR pressure to gas pressure is almost an order of magnitude smaller averaged over the scale height than at the midplane, due to the rapid loss of CRs with height when $\tau_{\rm abs}$ or $\tau_{\rm stream}$ are large. Conversely, at low optical depth and low $f_{\rm Edd}$, the ratio of CR pressure to gas pressure averaged over a scale height is generally \textit{larger} than it is at the midplane, due to the larger scale height of the CRs compared to the gas in these models.

\begin{figure}
    \centering
    \includegraphics[width=\columnwidth]{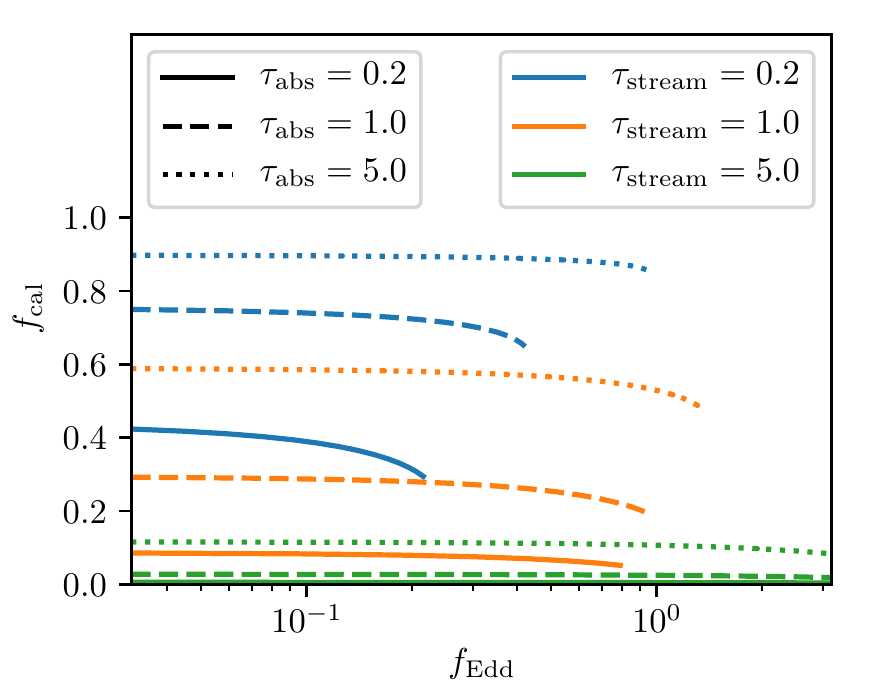}
    \caption{Fraction of CRs flux $f_{\rm cal}$ lost to  $pp$ collisions, and thus available to produce $\gamma$-ray emission, as a function of $f_{\rm Edd}$. We show solutions for a sample of absorption and streaming optical depths $\tau_{\rm abs}$ (solid, dashed, and dotted lines) and $\tau_{\rm stream}$ (blue, orange, and green lines), as indicated in the legend. The cases shown are the same as in \autoref{plotPressureRatioForCRpaper}.}
    \label{fig:CR_calorimetry}
\end{figure}

The primary observational signature of hadronic CRs beyond the Milky Way is $\gamma$-ray emission, and it is therefore interesting to ask what fraction of relativistic CRs are absorbed in collisions (i.e., lost to pion production) and thus are available to produce observable $\gamma$-rays. We can compute this calorimetric fraction from our solutions for the density and pressure profiles $r(\xi)$ and $p_c(\xi)$ in two ways. One is simply to note that the rate per unit volume at which CRs are lost to $pp$ collusions is $u_c/t_{\rm col}$. Thus if we integrate over the full gas column, and then divide by the flux $F_c(0)$ of CRs injected per unit area, we have
\begin{equation}
    f_{\rm cal} = \frac{1}{F_c(0)} \int_0^\infty \frac{u_c}{t_{\rm col}}.
\end{equation}
It is straightforward, if algebraically tedious, to rewrite the right hand side in terms of non-dimensional quantities using the transformations given in \autoref{ssec:non-dimensionalisation}. However, one can obtain the same result with significantly more insight by instead starting from the dimensionless CR transport equation, \autoref{eq:PcrSqrdDimless}. Let us define $q = -(ds/d\xi)^{-q} dp_c/d\xi = \mathcal{F}_c/K_* \beta$ as the dimensionless, scaled CR flux; from \autoref{BC_3}, we have $q(0) = f_{\rm Edd}$. If we now divide both sides of \autoref{eq:PcrSqrdDimless} by the injected CR flux $q(0)$ and then integrate from $0$ to $\xi$, the result is
\begin{equation}
    \frac{q(\xi)}{f_{\rm Edd}} = 1 - \frac{\tau_{\rm abs}}{f_{\rm Edd}} \int_0^\xi r p_c \, d\xi - \frac{\tau_{\rm stream}}{f_{\rm Edd}} \left[p_c(0) - p_c(\xi)\right].
\end{equation}
This expression has a simple physical interpretation: the quantity on the left hand side, $q(\xi)/f_{\rm Edd}$, is simply the fraction of the flux that was injected at $\xi=0$ that remains once the CRs have gotten to height $\xi$. The right hand side asserts that this fraction is equal to unity minus the flux that has been lost to absorption / pion losses (the term proportional to $\tau_{\rm abs}$) and to streaming losses (the term proportional to $\tau_{\rm stream}$). We can therefore identify the fraction of the flux that goes into pion production as
\begin{equation}
    \label{eq:f_abs}
    f_{\rm cal} = \frac{\tau_{\rm abs}}{f_{\rm Edd}} \int_0^\infty r p_c \, d\xi.
\end{equation}

In \autoref{fig:CR_calorimetry}, we show calculations of $f_{\rm cal}$ for the same set of models as shown in \autoref{plotPressureRatioForCRpaper}. Clearly the value of $f_{\rm Edd}$ has relatively little effect on $f_{\rm cal}$. Instead, the dominant parameters controlling $f_{\rm cal}$ are the streaming and absorption optical depths. If $\tau_{\rm abs} > \tau_{\rm stream}$ and $\tau_{\rm abs} > 1$, then a majority of the CRs are absorbed and can produce $\gamma$-ray emission. By contrast, if $\tau_{\rm stream} \gtrsim \tau_{\rm abs}$ or $\tau_{\rm abs} < 1$, then $f_{\rm cal}$ is much smaller.

\subsection{Model grid}
\label{ssec:model_grid}

Having developed some intuition for how the results of interest depend on the model parameters, we now generate a broad grid of solutions and extract pertinent parameters from them. Our grid consists of gas fractions $f_{\rm gas}$ from 0 to 1 in steps of 0.1, log Eddington ratios $\log f_{\rm Edd}$ from $10^{-4}$ to $10$ in steps of 0.025 dex, log absorption optical depths $\log \tau_{\rm abs}$ of $10^{-1.5}$ to $10^{2}$ in steps of 0.25 dex, and log streaming optical depths $\log\tau_{\rm stream}$ of $10^{-1}$ to $10^{1}$ in steps of 0.25 dex. Note that, for large enough $f_{\rm Edd}$, the model does not converge, and no equilibrium exists; we defer further discussion of this behaviour to \citetalias{Crocker2020b}. For each grid point where a solution is found, we record the midplane density and pressure, $r(0)$ and $p_c(0)$ and the fraction $f_{\rm cal}$ of the CR flux that is absorbed and therefore available for pion production (\autoref{eq:f_abs}).

\section{Implications for star-forming systems}
\label{sec:implications}

\subsection{Dimensionless parameters for observed systems}
\label{ssec:dimensionless_params}

We now have in hand machinery required to calculate the quantities of interest for any combination of dimensionless parameters. For a specified choice of CR propagation model, we also have in hand the mapping from galaxy gas surface density $\Sigma_{\rm gas}$, velocity dispersion $\sigma$, and gas fraction $f_{\rm gas}$, to the dimensionless optical depths $\tau_{\rm stream}$ and $\tau_{\rm abs}$ (\autoref{tab:CR_models}). The final dimensionless quantity we require is the Eddington ratio $f_{\rm Edd}$ (\autoref{BC_3}). This depends on the reference flux $F_*$ (\autoref{eq:defnFstar}) and on the injected CR flux $F_{c,0}$. We choose to write the latter in terms of the star formation rate, as
\begin{equation}
    F_{c,0} = \epsilon_{c,1/2} \dot{\Sigma}_\star,
\end{equation}
where $\epsilon_{\rm c,1/2}$ is the mean total energy in relativistic CRs released into each galactic hemisphere per unit mass of star formation. This yields
\begin{eqnarray}
    f_{\rm Edd} & = & \epsilon_{c,1/2}  \left(\frac{\tau_{\rm stream}}{\beta_s}\right) \frac{f_{\rm gas} \dot{\Sigma}_\star}{\pi G c \Sigma_{\rm gas}^2} \nonumber \\
    & = & 2.0 f_g K_* \dot{\Sigma}_{\star,-3} \sigma_1^{-1} \Sigma_{\rm gas,1}^{-2},
    \label{eq:fEdd_scaling}
\end{eqnarray}
where $\dot{\Sigma}_{\star,-3}=\dot{\Sigma}_\star/10^{-3}$ $M_\odot$ pc$^{-2}$ Myr$^{-1}$, and the numerical evaluation is for our fiducial value of the CR energy release per unit mass of stars formed, $\epsilon_{c,1/2}=\epsilon_{c,*,1/2}$ (see below). This equation contains a crucial result, which will become important later in our discussion: constant Eddington ratio corresponds roughly to $\dot{\Sigma}_\star \propto \Sigma_{\rm gas}^2$.

Accounting only for CR acceleration associated with core collapse supernovae, a reference value for the CR energy release per unit mass of star formation into each Galactic hemisphere, $\epsilon_{\rm c,1/2}$, can be defined as
\begin{eqnarray}
\epsilon_{\rm c,1/2} & \simeq & \frac{1}{2} \frac{\eta_{\rm c} \ E_{\rm SN}}{M_{\rm \star,SN}}  \nonumber \\
& \equiv & \epsilon_{\rm c,\star,1/2} \ \left(\frac{\eta_{\rm c}}{0.1}\right) \left(\frac{E_{\rm SN}}{10^{51} \ {\rm erg}}\right) \left(\frac{90 \ \msun}{M_{\rm \star,SN}}\right)
\end{eqnarray}
where $\eta_{\rm c} \sim 0.1$ is a rough \citep[e.g.,][]{Drury1989,Hillas2005,Strong2010,
Lacki2010,Paglione2012,Peng2016} calibration for the fraction of the total core collapse supernova kinetic energy release that ends up in CRs, $E_{\rm SN}$ is the supernova kinetic energy, and, for a \citet{Chabrier2005} IMF, one core collapse supernova requires the formation of $M_{\rm \star,SN} \simeq 90 \ \msun$ of stars assuming that all stars born with masses of 8 $\msun$ or above end their lives as core collapse supernovae. Numerically, the normalising cosmic ray efficiency is
\begin{equation}
\epsilon_{\rm c,\star,1/2} \equiv 5.6 \times 10^{47} \ {\rm erg} \ \msun^{-1}  
\, .
\label{eq:epsStar}
\end{equation}
Note that this normalisation for  $\epsilon_{\rm c,1/2}$ may be too conservative as it ignores other sources of mechanical power that may end up in CRs including stellar wind shocks, pulsar winds, and thermonuclear supernovae. It also neglects the possibility, for which there is some evidence \citep{Nomoto2006}, that the mean mechanical energy per core collapse supernova might exceed by a factor of a few the canonical $10^{51}$ erg, and that a fraction of massive star core collapses end in black hole formation with potentially much weaker supernovae, or none at all \citep[e.g.,][]{Pejcha2015, Gerke2015}. Finally, this normalisation neglects the possibility that some fraction of CRs produced may be trapped in the SN-driven hot phase of the ISM and then advected out of the galaxy in a galactic wind while having relatively little interaction with the neutral phase; CRs that follow this path contribute to neither pressure support nor $\gamma$-ray emission, and thus advective escape of hot gas might lower the effective value of $\epsilon_{c,1/2}.$ (Advective escape of neutral gas is likely unimportant, since, even in the galaxies with the strongest winds, only a small fraction of the neutral material is ejected per dynamical time.) We will ignore these complications in the remainder of this paper, however.

\begin{figure}
    \centering
    \includegraphics[width=\columnwidth]{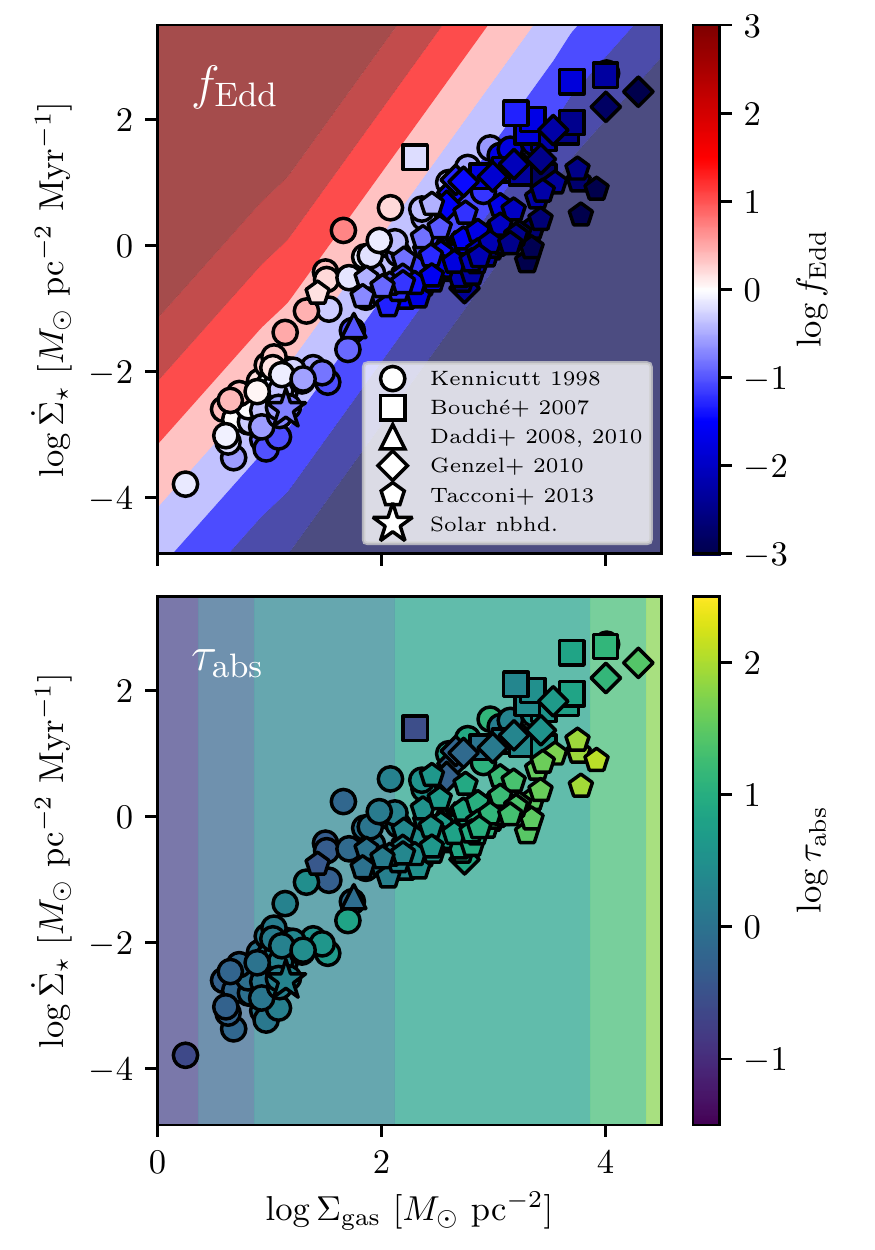}
    \caption{Distribution of $f_{\rm Edd}$ (top panel) and $\tau_{\rm abs}$ (bottom panel) values for a sample of observed galaxies culled from the literature, computed using the ``Streaming'' CR transport model. Points are coloured by the value of $f_{\rm Edd}$ or $\tau_{\rm abs}$ that we infer for that galaxy, following the discussion in the main text; colour bars indicate numerical values, and the shape of the symbol indicates the sample from which it is drawn: \citet{Kennicutt1998}, \citet{Bouche2007}, \citet{Daddi2008, Daddi2010a}, \citet{Genzel2010}, or \citet{Tacconi2013}; the star indicates Solar neighbourhood conditions, for which we adopt values described in the main text. Coloured contours indicate regions of the plane with values of $\log f_{\rm Edd}$ from $-3$ to $3$ in steps of 1 and $\log \tau_{\rm abs}$ from $-1.5$ to $2.5$ in steps of 0.5, respectively, for the streaming CR propagation model \autoref{sssec:FLRW}, using gas fractions, velocity dispersions, and ionisation fractions interpolated as a function of $\Sigma_{\rm gas}$ as described in the main text.}
    \label{fig:ks_plane_parameters}
\end{figure}

\begin{figure}
    \centering
    \includegraphics[width=\columnwidth]{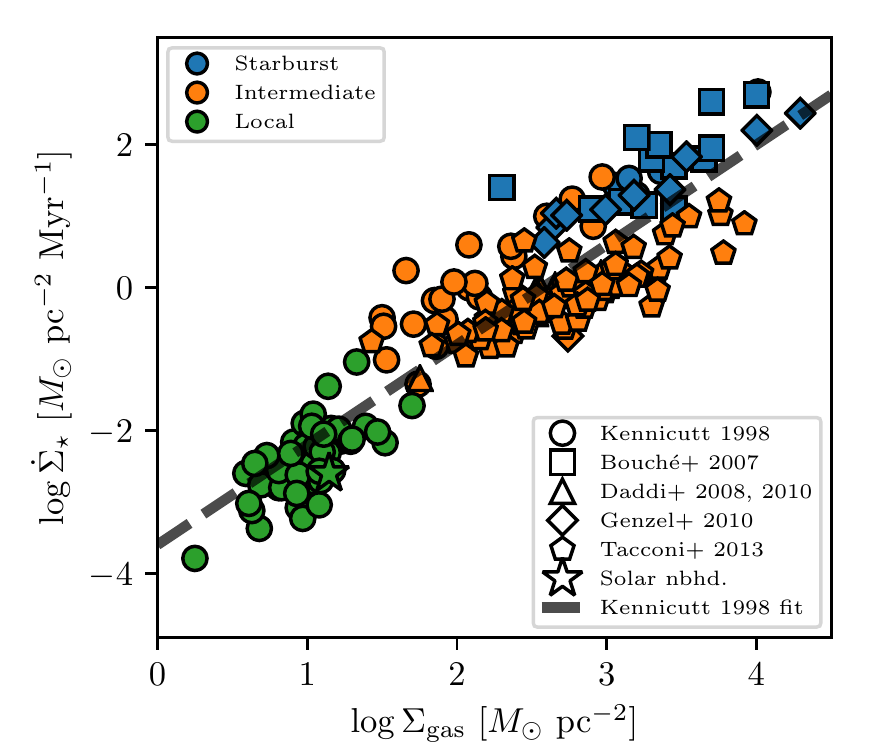}
    \caption{Illustration of our classification of galaxies in the Kennicutt-Schmidt plane as ``local'', ``intermediate'', and ``starburst''. Colour indicates the classification, while symbol shape indicates the sample from which the galaxy is drawn. Points match those shown in \autoref{fig:ks_plane_parameters}. For reference, we also overlay (dashed black line) the \citet{Kennicutt1998} fit for the relationship between star formation and gas surface density.}
    \label{fig:ks_plane_classifications}
\end{figure}

We show values of $f_{\rm Edd}$ and $\tau_{\rm abs}$ for a sample of galaxies culled from the literature in \autoref{fig:ks_plane_parameters}, computed adopting the ``Streaming'' model for CR transport (\autoref{sssec:FLRW}) with $M_A=1.5$; we do not show $\tau_{\rm stream}$, since for the ``Streaming'' model it is simply $M_A^3$, and thus is constant for all galaxies. We compare to the ``Scattering'' model, and explore the dependence on $M_A$, in \autoref{ssec:transport_models}. The data come from the compilation of \citet{Krumholz2012b}, and consist of measurements of gas surface density $\Sigma_{\rm gas}$ and star formation rate surface density $\dot{\Sigma}_\star$. We also add a point to represent conditions in the Solar neighbourhood, which has $\Sigma_{\rm gas} \approx 14$ $M_\odot$ pc$^{-2}$ \citep{McKee2015} and $\dot{\Sigma}_\star\approx 2.5\times 10^{-3}$ $M_\odot$ pc$^{-2}$ Myr$^{-1}$ \citep{Fuchs2009}. Since velocity dispersions and gas fractions are only available in the literature for a small subset of these galaxies, we assign values as follows: for the sample of \citet{Kennicutt1998}, we adopt ``Local'' parameters for all galaxies classified as spirals by \citeauthor{Kennicutt1998}, and we also adopt these properties for the Solar neighbourhood; for those classified as starburst, we adopt ``Intermediate'' parameters if the gas surface density is below $10^3$ $M_\odot$ pc$^{-2}$, and ``Starburst'' parameters otherwise. We adopt ``Intermediate'' parameters for the entire sample of galaxies from \citet{Daddi2008}, \citet{Daddi2010a}, and \citet{Tacconi2013}, and for all galaxies from the sample of \citet{Genzel2010} except those that \citeauthor{Genzel2010} classify as sub-mm galaxies, for which we use ``Starburst'' parameters. Finally, we also apply ``Starburst'' parameters for the sample of sub-mm galaxies taken from \citet{Bouche2007}. We illustrate the classifications in \autoref{fig:ks_plane_classifications}; for reference, we also overlay on this figure the \citet{Kennicutt1998} fit for the relationship between star formation and gas surface densities.\footnote{Note that the \citet{Kennicutt1998} line shown in \autoref{fig:ks_plane_classifications} does not in fact pass through the data points in the \citet{Kennicutt1998} sample. This is because the data points have been adjusted to use updated estimates of the conversion from CO luminosity to gas surface density, and from infrared or H$\alpha$ luminosity to star formation surface density, following \citet{Daddi2010a}. However, we choose not to adjust the fit for these updates, in part to maintain consistency with earlier work, and in part because the fit remains a reasonable one for the expanded data set shown in the figure.} Similarly, in order to overlay rough contours on \autoref{fig:ks_plane_parameters}, we linearly interpolate $\log\sigma$, $\log f_{\rm gas}$, and $\log \chi$ as a function of $\log\Sigma_{\rm gas}$ between the three cases listed in \autoref{tab:galparams}, treating each case as a single point at the center of the stated range. However, we emphasise that all these classifications and parameter choices are approximate. More accurate estimates would use values of the gas fraction and velocity dispersion determined galaxy-by-galaxy, and estimates of $\chi$ based on detailed chemical modelling (c.f.~\citealt{Krumholz2019}).

The primary conclusion to be drawn from the figure is that, as one proceeds along the star-forming galaxy sequence from low to high gas and star formation surface density, galaxies become increasingly sub-Eddington and optically thick to CRs. Local spirals and dwarfs tend to have $f_{\rm Edd}\sim 0.1 - 1$ and $\tau_{\rm abs} \ll 1$, while high-redshift galaxies and starbursts typically have $f_{\rm Edd} \sim 0.001 - 0.1$ and $\tau_{\rm abs} \sim 1-10$.

\subsection{CR pressures}
\label{ssec:cr_pressures}

\begin{figure}
    \centering
    \includegraphics[width=\columnwidth]{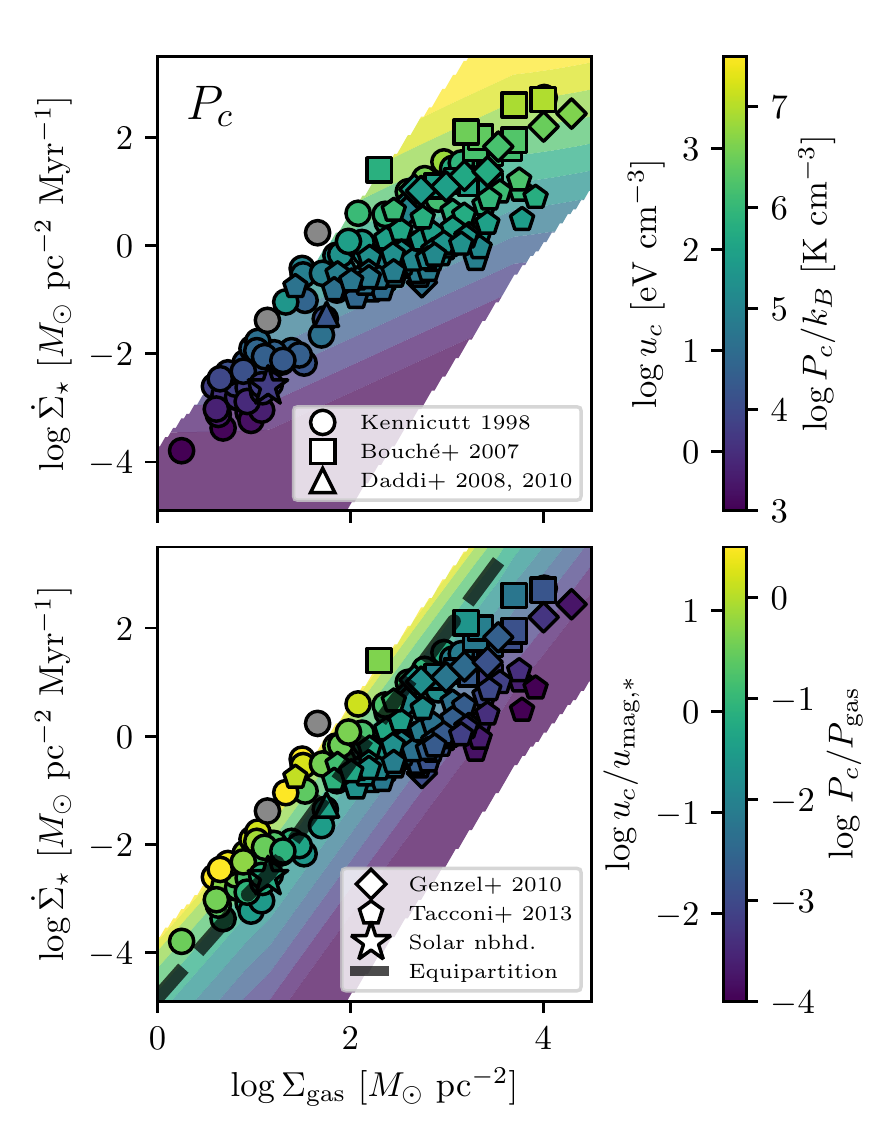}
    \caption{Estimated CR pressure and energy density (top panel) and ratio of CR pressure to gas pressure (bottom panel) at the galactic midplane for the sample of observed galaxies shown in \autoref{fig:ks_plane_parameters} (coloured points), computed using the ``Streaming'' CR transport model. Grey points mark galaxies whose Eddington ratios place them outside our grid. We also show contours of $P_c$ and $P_c/P_{\rm gas}$, computed by interpolating as in \autoref{fig:ks_plane_parameters}. The contours of $P_c$ run from $P_c/k_B = 10^3 - 10^{7.5}$ K cm$^{-3}$ in steps of 0.5 dex, and the contours of $P_c/P_{\rm gas}$ run from $10^{-4} - 10^{0.5}$ in steps of 0.5 dex. Points that are not covered by contours correspond to combinations of parameters $f_{\rm g}$, $f_{\rm Edd}$, $\tau_{\rm abs}$ and $\tau_{\rm stream}$ that are outside our grid of solutions. The black dashed line corresponds to the locus of equality between CR and magnetic energy densities.}
    \label{fig:ks_plane_pressure}
\end{figure}

We show estimates for the midplane CR pressure, and the ratio of CR pressure to gas pressure, in \autoref{fig:ks_plane_pressure}; results for the average pressure over the first scale height are qualitatively similar. In order to generate these plots, for each galaxy we compute $\log f_{\rm Edd}$, $\log \tau_{\rm abs}$, $\log \tau_{\rm stream}$, and $f_{\rm gas}$ as described in \autoref{ssec:dimensionless_params}, and then linearly interpolate on our grid of solutions (\autoref{ssec:model_grid}) to produce predicted values of $\log p_c$ and $\log r = \log (ds/d\xi)$.\footnote{A few galaxies, indicated by the grey points in \autoref{fig:ks_plane_pressure}, fall outside our grid, at values of $f_{\rm Edd}$ too high for a solution to exist. We discuss the significance of the maximum value of $f_{\rm Edd}$ in \citetalias{Crocker2020b}, and here simply note that, while the best estimates for these galaxies' properties are off our grid, they are off by only a very small amount, and any plausible estimate of the errors bars (at least a factor of two in both directions, likely more) overlaps the grid extensively.} We then scale these back from dimensionless to physical units using the transformations given in \autoref{ssec:non-dimensionalisation}. Similarly, we generate the contours in the background using the same interpolation scheme as described in \autoref{ssec:dimensionless_params}.

We see that typical midplane CR pressures range from $P_c/k_B \sim 10^{3.5}$ K cm$^{-3}$ (energy density $u_c \sim 1$ eV cm$^{-3}$) for sub-Milky Way galaxies up to $\sim 10^7$ K cm$^{-3}$ (energy density $\sim \mathrm{few}$ keV cm$^{-3}$) for the most intensely star-forming galaxies. Not surprisingly, midplane CR pressure increases with star formation rate. However, we also see that the ratio of CR pressure to gas pressure decreases systematically with star formation rate, such that $P_{\rm c}/P_{\rm gas}$ is typically $\sim 0.1-1$ for galaxies with $\Sigma_{\rm gas} \lesssim 100$ $M_\odot$ pc$^{-2}$, but drops to $\sim 10^{-3}$ for galaxies with $\Sigma_{\rm gas} \gtrsim 1000$ $M_\odot$ pc$^{-2}$. Contours of constant $P_c/P_{\rm gas}$ are close to lines of slope 2 in the lower panel of \autoref{fig:ks_plane_pressure} (i.e., $\dot{\Sigma}_\star \propto \Sigma_{\rm gas}^2$), whereas the observed distribution of galaxies forms a significantly shallower relationship. Thus we find that CRs are dynamically significant for weakly star-forming, low surface density galaxies, but become increasingly-unimportant as we move to higher surface density, more strongly star-forming galaxies.

It is worth pointing out that, although we are comparing CR pressure to gas pressure in \autoref{fig:ks_plane_pressure}, we can also read the figure as describing the ratio of CR and magnetic energy densities, and thus the extent to which equipartition between CRs and magnetic fields holds. Defining the midplane magnetic energy density $u_{\rm mag,*}=B_*^2/8\pi$, and making use of \autoref{eq:Bstar}, we can write the ratio of CR to magnetic energy density at the midplane as
\begin{equation}
    \frac{u_c}{u_{\rm mag,*}} = 6 M_A^2 \frac{P_c}{P_{\rm gas}}.
\end{equation}
Thus for our fiducial choice $M_A=1.5$, equipartition between CRs and magnetic fields corresponds to $P_c/P_{\rm gas}\simeq 0.1$. Thus \autoref{fig:ks_plane_pressure} can be read as also giving $u_c/u_{\rm mag,*}$, if we simply shift the colour scale up by $\simeq 1$ dex, i.e., $\log (P_c/P_{\rm gas}) \simeq -1$ corresponds to $u_c/u_{\rm mag,*}\simeq 1$. We show the locus $u_c/u_{\rm mag,*} = 1$ as the black dashed line in the lower panel of \autoref{fig:ks_plane_pressure}. We see that the Solar neighbourhood, and galaxies with similar conditions, are expected to show near-equipartition between CRs and magnetic fields. However, as we move to galaxies that are forming stars within increasing vigour, to the right of \autoref{fig:ks_plane_pressure}, CRs fall below equipartition with the magnetic field by $1-2$ orders of magnitude \citep{Thompson2006,Lacki2010,Lacki_Beck}.

\begin{figure}
    \centering
    \includegraphics[width=\columnwidth]{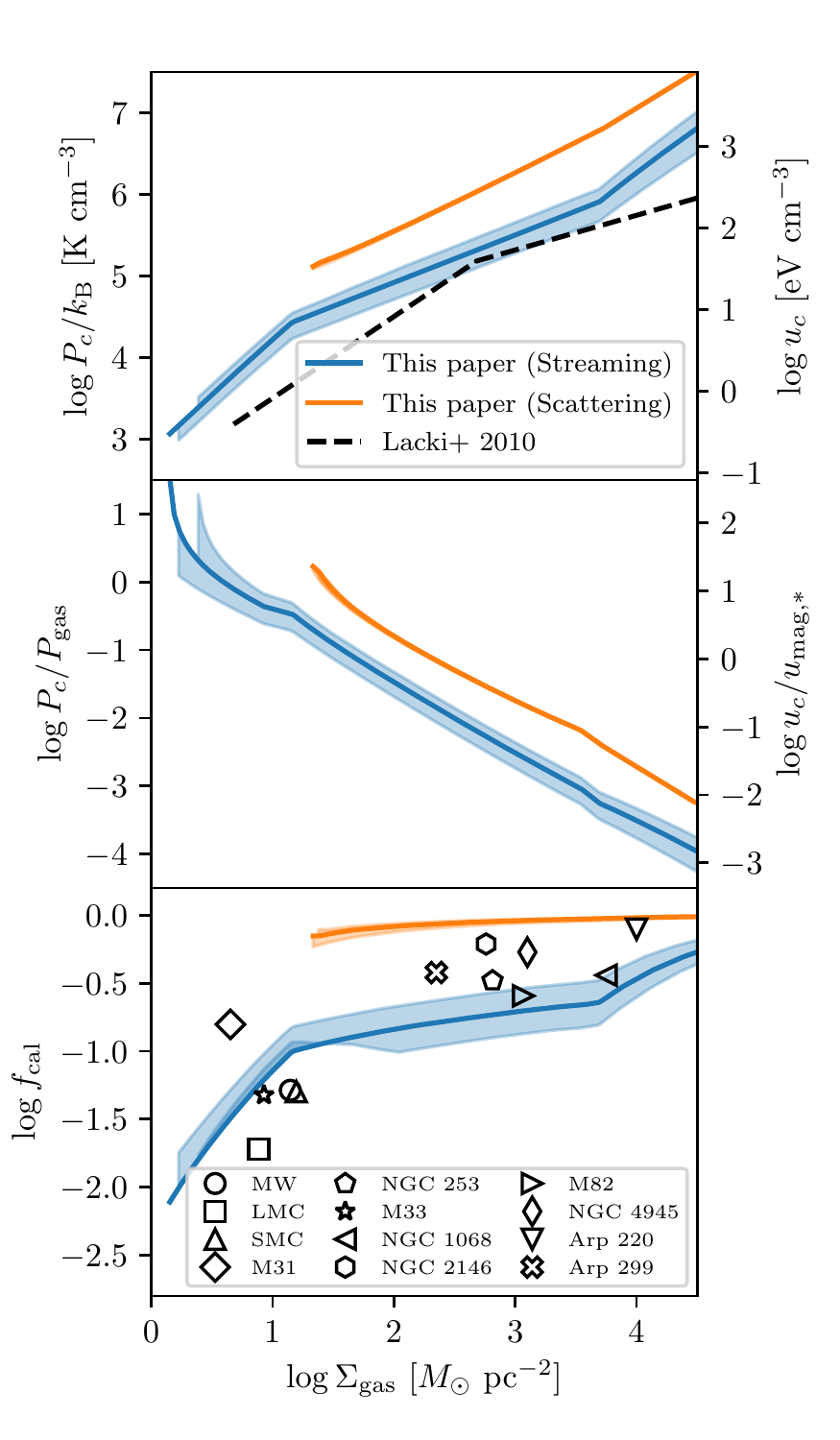}
    \caption{CR pressure and energy density (top panel), ratio of CR pressure to gas pressure (middle panel), and calorimetry fraction (bottom panel) computed as a function of gas surface density, taking the star formation surface density to be the mean value given by the \citet{Kennicutt1998} relation, as illustrated in \autoref{fig:ks_plane_classifications}. We obtain ancillary data properties ($\sigma$, $f_{\rm gas}$, $\chi$) along this line by interpolating, using the same procedure as is used to construct the contours in \autoref{fig:ks_plane_parameters}. The blue and orange curves indicate the results for streaming and scattering CR transport models respectively, with the central solid line indicating the reult for our fiducial Alfv\'en Mach number $M_A = 1.5$, and the shaded enclosing region showing the results for $M_A = 1 - 2$. Note that $f_{\rm Edd}$ generally increases toward lower surface density as one moves along the \citet{Kennicutt1998} relation, and, as a result, for each of the transport models shown there is a minimum surface density below which $f_{\rm Edd}$ is large enough that we can no longer find a hydrostatic solution. The model curves terminate at this surface density.}
    \label{fig:ks_plane_cut_pressure}
\end{figure}

It is worth noting that our conclusion that CR pressure is smaller than gas pressure in starbursts, and that the CR energy density is sub-equipartition, is consistent with the one-zone models developed by \citet{Lacki2010} to study the far infrared-radio correlation. We illustrate this in the top two panels of \autoref{fig:ks_plane_cut_pressure}, where we show our estimated CR pressure and ratio of CR to gas pressure computed along the \citet{Kennicutt1998} relation. That is, the figure is a parametric plot showing the values indicated by the contours in \autoref{fig:ks_plane_pressure}, calculated along a path through the $\Sigma_{\rm gas}-\dot{\Sigma}_\star$ plane given by the \citeauthor{Kennicutt1998} fit, and illustrated in \autoref{fig:ks_plane_classifications}. 
In the top panel, we compare our estimated $P_c$ values  to those obtained by \citet{Lacki2010}.
Clearly the results are qualitatively similar,
with the ``Scattering" curve somewhat closer to 
\citeauthor{Lacki2010}'s results for our fiducial parameter choices. Note that \citeauthor{Lacki2010}'s calculations were empirically constrained to reproduce the observed far infrared-radio correlation and were used to predict $\gamma$-ray fluxes and calorimetric fractions from star-forming galaxies across the \cite{Kennicutt1998} relation. A crucial point of this analysis involves not just the ratio of CR to magnetic energy densities, but also the ratio of magnetic energy density to photon energy density, which controls the relative importance of synchrotron and inverse Compton cooling for CR electrons. We will explore the predictions of our models for emission from CR electrons in a future paper in this series.

\subsection{CR calorimetry}

\begin{figure}
    \centering
    \includegraphics[width=\columnwidth]{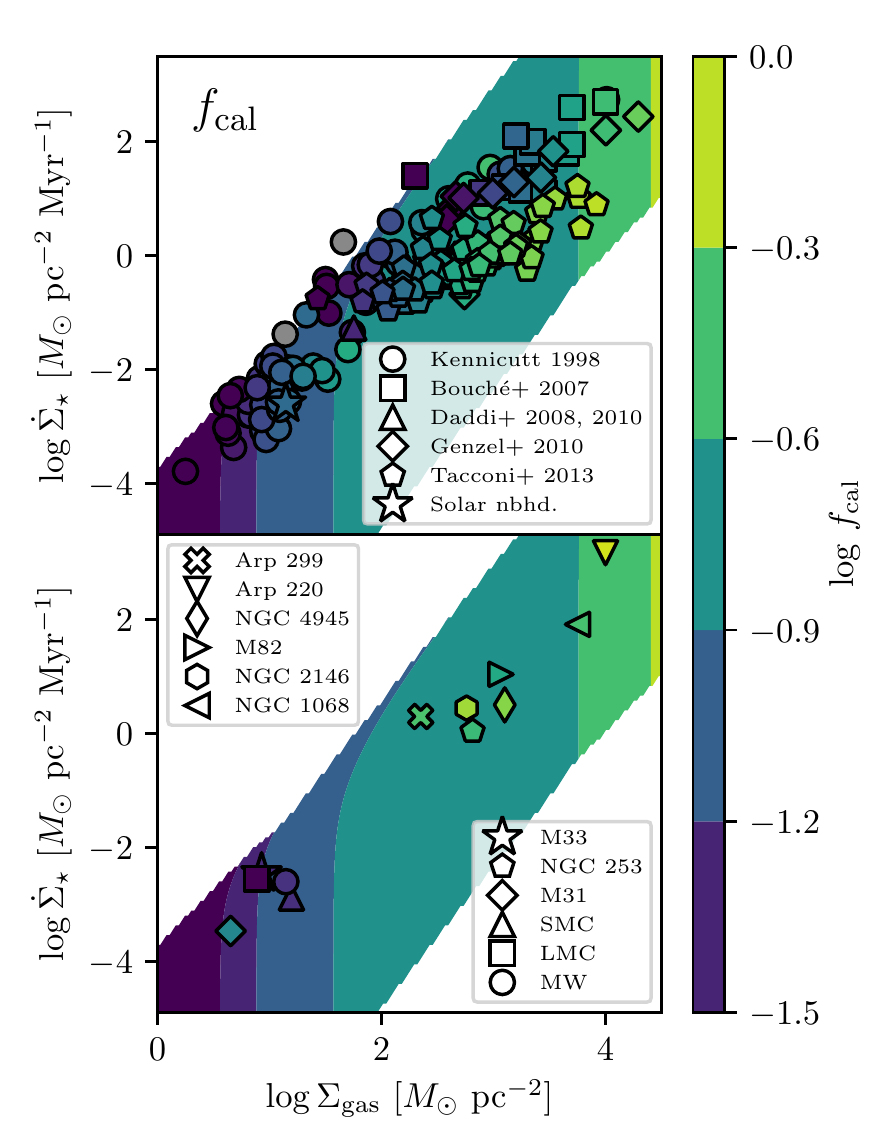}
    \caption{Fraction of CR flux that is absorbed, and thus available to produce $\gamma$-rays ($f_{\rm cal}$). In the top panel we show this quantity estimated for the sample of observed galaxies shown in \autoref{fig:ks_plane_parameters} (coloured points), computed using the ``Streaming'' CR transport model. Grey points mark galaxies whose Eddington ratios place them outside our grid. We also show contours of $\log f_{\rm cal}$, running from $-1.5$ to 0 in steps of $0.3$, interpolated across the plane using the same method as used in \autoref{fig:ks_plane_parameters}. Points that are not covered by contours correspond to combinations of parameters $f_{\rm g}$, $f_{\rm Edd}$, $\tau_{\rm abs}$ and $\tau_{\rm stream}$ that are outside our grid of solutions. In the bottom panel, we show the same background contours, but the data points and their colours now indicate gas surface densities, star formation rates, and  observationally-estimated calorimetry fractions for the galaxies listed in \autoref{tab:calorimetry}.}
    \label{fig:ks_plane_calorimetry}
\end{figure}

\begin{table*}
    \centering
\begin{tabular}{lccccccccc}
\hline\hline
Galaxy  &  $\log\Sigma_{\rm gas}$  &  $\log\dot{\Sigma}_\star$            &  Type  &  $\log \dot{M}_\star$  &  $\log L_\gamma$  &  $\log f_{\rm cal,obs}$  &  $\log f_{\rm cal,str}$  &  $\log f_{\rm cal,sca}$ \\
        &  [$M_\odot$ pc$^{-2}$]   &  [$M_\odot$ pc$^{-2}$ Myr$^{-1}$]  &        &  [$M_\odot$ yr$^{-1}$]         &  [erg s$^{-1}$] \\ \hline
Milky Way (MW)  &  $ 1.15$   &  $-2.60$  &  Local  &  $\phantom{-} 0.28$  &  $38.91$  &  $-1.29$  &  $-0.90$  &  $-0.10$ \\
LMC  &  $ 0.89$   &  $-2.55$  &  Local  &  $-0.70$  &  $37.50$  &  $-1.72$  &  $-1.15$  &  $\ldots$ \\
SMC  &  $ 1.20$   &  $-2.89$  &  Local  &  $-1.48$  &  $37.14$  &  $-1.30$  &  $-0.85$  &  $-0.09$ \\
NGC 224 (M31)  &  $ 0.65$   &  $-3.47$  &  Local  &  $-0.46$  &  $38.66$  &  $-0.80$  &  $-1.33$  &  $-0.23$ \\
NGC 253  &  $ 2.81$   &  $\phantom{-} 0.04$  &  Intermediate  &  $\phantom{-} 0.61$  &  $40.05$  &  $-0.48$  &  $-0.67$  &  $-0.03$ \\
NGC 598 (M33)  &  $ 0.93$   &  $-2.46$  &  Local  &  $-0.35$  &  $38.25$  &  $-1.32$  &  $-1.11$  &  $\ldots$ \\
NGC 1068  &  $ 3.75$   &  $\phantom{-} 1.92$  &  Starburst  &  $\phantom{-} 1.44$  &  $40.92$  &  $-0.44$  &  $-0.77$  &  $-0.02$ \\
NGC 2146  &  $ 2.76$   &  $\phantom{-} 0.45$  &  Intermediate  &  $\phantom{-} 1.24$  &  $40.95$  &  $-0.21$  &  $-0.70$  &  $-0.03$ \\
NGC 3034 (M82)  &  $ 3.07$   &  $\phantom{-} 1.04$  &  Intermediate  &  $\phantom{-} 0.94$  &  $40.27$  &  $-0.59$  &  $-0.52$  &  $-0.02$ \\
NGC 4945  &  $ 3.10$   &  $\phantom{-} 0.51$  &  Intermediate  &  $\phantom{-} 0.65$  &  $40.30$  &  $-0.27$  &  $-0.50$  &  $-0.02$ \\
Arp 220  &  $ 4.00$   &  $\phantom{-} 3.18$  &  Starburst  &  $\phantom{-} 2.38$  &  $42.20$  &  $-0.10$  &  $-0.61$  &  $-0.01$ \\
Arp 299  &  $ 2.35$   &  $\phantom{-} 0.30$  &  Intermediate  &  $\phantom{-} 2.05$  &  $41.55$  &  $-0.42$  &  $-1.02$  &  $-0.06$ \\
\hline
\end{tabular}
    \caption{Observed and theoretically-estimated calorimetry fractions for a sample of \textit{Fermi}-detected galaxies. Columns are as follows: (1) galaxy name; (2) gas surface density; (3) star formation surface density; (4) classification as local (Loc), intermediate (Int), or starburst (SB); (5) star formation rate; (6) $\gamma$-ray luminosity; (7) observationally-estimated calorimetric fraction, computed from \autoref{eq:fcal_obs}; (8) theoretical estimate of $f_{\rm cal}$, computed from \autoref{eq:f_abs}, assuming the streaming CR transport model; (9) same as column (8), but using the scattering transport model; an entry of $\ldots$ indicates that the estimated parameters for this galaxy place it outside our model grid. Data sources: all $\gamma$-ray luminosities $L_\gamma$ are taken from \citet{Ajello2020}, except for those for Arp 220 (from \citealt{Griffin16a}) and the Milky Way (from \citealt{Ackermann2012}). All SFRs for objects classified as Intermediate or Starburst are obtained by converting the total IR luminosity given by \citet{Ajello2020} to a SFR using the conversion given in Table 1 of \citet{Kennicutt2012}. Remaining gas and SFR data are from the following sources: Milky Way -- gas surface density from \citet{McKee2015}, SFR surface density from \citet{Fuchs2009}, total SFR from \citet{Chomiuk2011}; LMC and SMC -- total gas mass and SFR from \citet{Jameson2016}, values per unit area derived by dividing by an area $\pi R_{25}^2$, where we take $R_{25}$ from \citet{deVaucouleurs1991}; NGC 224 -- total SFR from \citet{Rahmani2016}, gas mass obtained by adding the H~\textsc{i} mass from \citet{Chemin2009} and the H$_2$ mass  from \citet{Nieten2006}, converted to areal quantities using a radius of 18 kpc from \citet{Kennicutt1998}; NGC 253, NGC 1068, NGC 2146, NGC 3035, NGC 4945, Arp 299 -- gas and SFR surface densities taken from \citet{Liu2015}, using gas values for their continuously-variable $\alpha_{\rm CO}$ case, and SFR values derived from IR; Arp 220 --  \citet{Kennicutt1998}, with  gas mass and SFR per unit area adjusted to use a conversion factor $\alpha_{\rm CO} = 0.8$ $M_\odot$ pc$^{-2}$ [K km s$^{-1}$]$^{-1}$, and to a \citet{Chabrier2005} IMF.} 
    \label{tab:calorimetry}
\end{table*}

We next examine the fraction $f_{\rm cal}$ of CRs that are lost to pion-producing collisions in \autoref{fig:ks_plane_calorimetry}. In the top panel, we show predicted calorimetry fractions for the same sample of galaxies plotted in \autoref{fig:ks_plane_parameters}. Here we see a trend that is generally the opposite of that in \autoref{fig:ks_plane_pressure}: local galaxies tend to have relatively low values of $f_{\rm cal}$, while higher surface density galaxies have higher values. Typical values in galaxies similar to the Milky Way are $\sim 5-10\%$, while the fraction rises to $\sim 50\%$ in galaxies at the top end of the star-forming sequence. 

At first one might be surprised that the difference in calorimetry across the star-forming sequence is as small as it is -- after all, the gas surface density increases by $\sim 4$ dex from the left to the right side of \autoref{fig:ks_plane_calorimetry}, so one might expect a similar level of variation in $f_{\rm cal}$. The main reason that the true variation is not so large, at least in the streaming model, is that the increase in surface density is partly countered by variations in the ionisation fraction $\chi$, which change the absorption optical depth as $\tau_{\rm abs} \propto \sqrt{\chi}$ (c.f.~\autoref{tab:CR_models}); the ionisation fraction is lower in the neutral ISM of starbursts than in local spirals due to their much high densities and thus recombination rates \citep{Krumholz2019}. Indeed, \citeauthor{Krumholz2019} show that this variation is critical to explaining the observed break in the $\gamma$-ray spectra of nearby starbursts above $\sim 1$ TeV. The ionisation fraction matters as a direct result of the dependence of the CR streaming speed on the Alfv\'en Mach number of the ions in a medium where ions and neutrals are decoupled: the lower the ionisation fraction, the faster the CR propagation speed and the less time it takes CRs to escape. This effect partially cancels out the increase in gas surface density, going from local spirals to starbursts, which is why $f_{\rm cal}$ rises only by a factor of $\sim 5-10$ across the star-forming sequence. 

In the bottom panel of \autoref{fig:ks_plane_calorimetry}, the background contours show the same predicted theoretical trend as in the upper panel, but now we overplot data points with colours for galaxies with \textit{Fermi}-detected $\gamma$-ray emission, for which it is possible to estimate the calorimetry fraction directly. Thus the data points in the upper panel of \autoref{fig:ks_plane_calorimetry} shows \textit{predicted} calorimetry fractions, while those in the lower panel show \textit{measured} (at least approximately) values. We derive our measured values from the observed star formation rate $\dot{M}_\star$ and $\gamma$-ray luminosity $L_{\gamma}$; we take the latter primarily from from \citet{Ajello2020}, and the former from a variety of sources in the literature as detailed in \autoref{tab:calorimetry}. We estimate the observed calorimetry fraction from these two as
\begin{equation}
\label{eq:fcal_obs}
f_{\rm cal,obs} = \frac{L_{\gamma}}{\zeta_{\rm CR} \dot{M}_\star}, 
\end{equation}
where $\zeta_{\rm CR} = 8.3\times 10^{39}$ erg s$^{-1}$ / ($M_\odot$ yr$^{-1}$). We derive the conversion factor $\zeta_{\rm CR}$ using from \citet[][their equation 11]{Lacki2011}, and assuming that (1) a fraction $\beta_\gamma=1/3$ of CRs with energies above the pion production threshold produce neutral pions that decay into $\gamma$-rays, (2) a fraction $\beta_\pi=0.7$ of the energy from these decays goes into $\gamma$-rays with energies high enough to be detected by \textit{Fermi} and thus contribute the measured $L_\gamma$, and (3) there is one supernova per $M_{\rm \star,SN}=90$ $M_\odot$ of stars formed, each of which explodes with total energy $10^{51}$ erg, of which a fraction $\eta_c=0.1$ does into CRs with energies $\geq 1$ GeV. The value of $\zeta_{\rm CR}$ should be regarded as uncertain that the factor of $\sim 2$ level. We list our derived values of $f_{\rm cal,obs}$ in \autoref{tab:calorimetry} (uncertain by a factor of $\sim 2$ due to uncertainties in $\zeta_{\rm CR}$), and values predicted by \autoref{eq:f_abs} for both the streaming and scattering transport models; for the purposes of this computation, we use the surface densities and star formation rates listed in the table, and classify galaxies as Local, Intermediate, and Starburst following the same scheme described in \autoref{ssec:dimensionless_params}. We defer a discussion of the scattering models to \autoref{ssec:transport_models}, but for now we note that, within the uncertainties in the calorimetric fraction, our streaming model provides reasonable agreement (within $\approx 0.5$ dex) for most galaxies. The largest discrepancies are with the brightest starbursts, where the observationally-estimated values of $f_{\rm cal}$ are in the range $\sim 50-80\%$, while our model tends to predict values a factor of $\sim 2$ smaller as a result of streaming losses. We can also see this effect in the bottom panel of \autoref{fig:ks_plane_cut_pressure}, where we show our predicted calorimetry fractions along the \citet{Kennicutt1998} relation. Our models provide reasonably good agreement for normal galaxies, but tend to underestimate the calorimetry fractions of starbursts by factors of $\sim 2$.

However, we note that there are a number of confounding factors that should be considered, In addition to the uncertainty on $\xi_{\rm CR}$, our observational estimates of the calorimetry fraction do not account for possible contributions to $L_\gamma$ from buried active galactic nuclei (AGN; possibly important in starbursts, though this seems unlikely to be a large effect in our sample, for which star formation dominates the bolometric output), and from non-hadronic processes (e.g., bremsstrahlung and inverse Compton emission) or millisecond pulsars (possibly important in galaxies with low star formation rates). Similarly, our theoretical models do not account for possible advective escape of CRs that are trapped in the hot phase of the ISM, and never interact with neutral gas. If these are significant, this would reduce $f_{\rm cal}$.

\subsection{Dependence on the CR transport model}
\label{ssec:transport_models}

\begin{figure*}
    \centering
    \includegraphics[width=\textwidth]{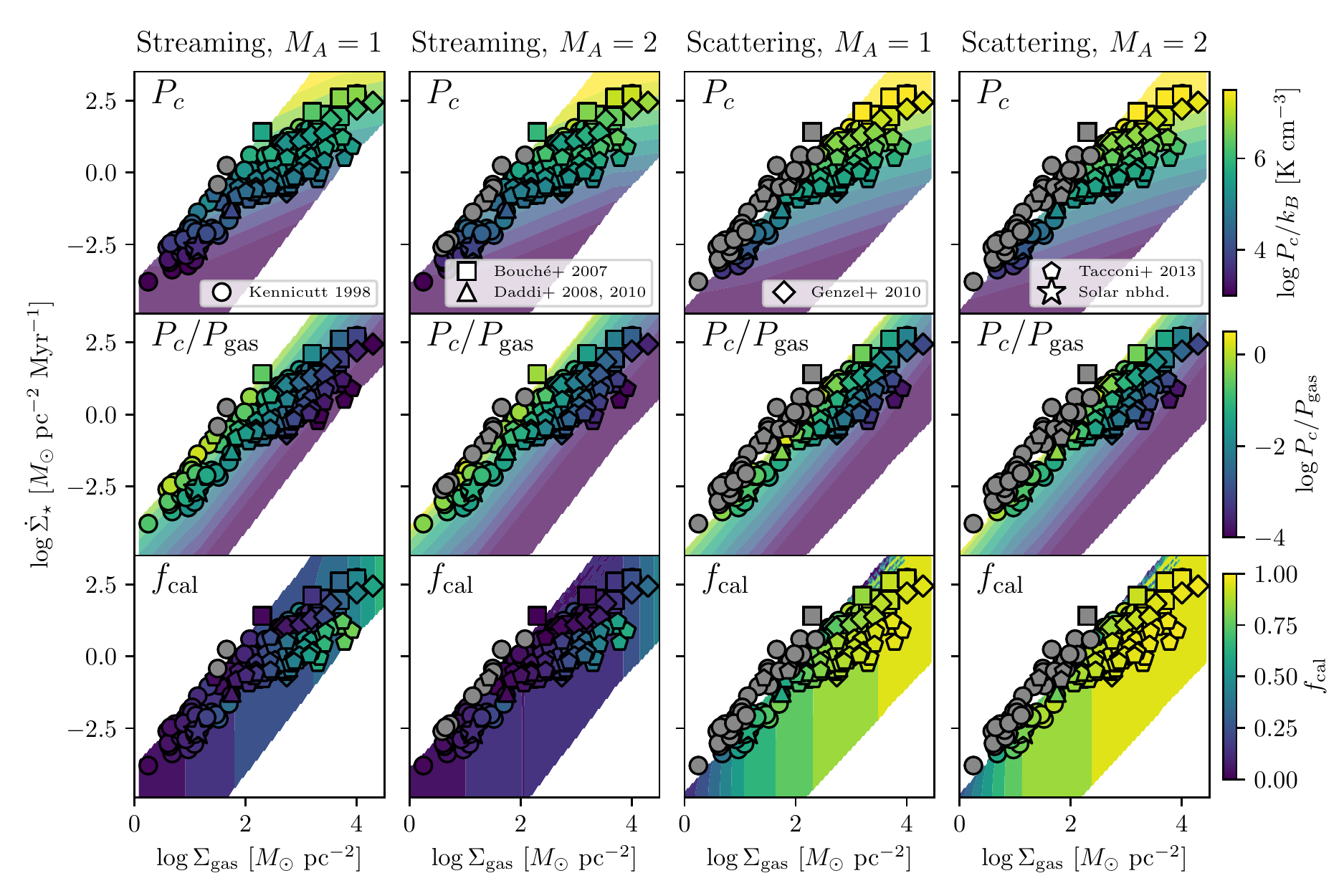}
    \caption{Comparison of results for different CR transport models. The left two columns show the ``Streaming'' model, computed using Alfv\'en Mach numbers $M_A=1$ and 2, compared to our fiducial choice $M_A=1.5$; the right two columns show the ``Scattering'' model for $M_A=1$ and 2. The top two rows show the midplane CR pressure $P_c$ and ratio of CR pressure to gas pressure $P_c/P_{\rm gas}$, and can be compared directly to \autoref{fig:ks_plane_pressure}; as in that figure, contours are in steps of $0.5$ dex, starting from a minimum of $P_c/k_B=10^3$ K cm$^{-3}$ and $P_c/P_{\rm gas} = 10^{-4}$. The bottom row shows the calorimetric fraction $f_{\rm cal}$, and is comparable to \autoref{fig:ks_plane_calorimetry}; however, note that, to avoid saturation, we use a different colour scale for $f_{\rm cal}$ here than we do in \autoref{fig:ks_plane_calorimetry}. Here contours run from $0-1$ in $f_{\rm cal}$, in steps of $0.1$.}
    \label{fig:ks_plane_crmodels}
\end{figure*}

We now turn to the question of how our results depend on our choice of CR transport model, and on parameters within that model. In \autoref{fig:ks_plane_crmodels} we show our computed CR pressures, ratios of CR to gas pressure, and calorimetric fractions for four different CR transport models: ``Streaming'' using $M_A=1$ and 2, and ``Scattering'' also using $M_A=1$ and 2. All other aspects of the calculation are identical to those discussed previously.

First examining the top two rows of \autoref{fig:ks_plane_crmodels}, we see that neither the value of $M_A$ nor the choice of CR transport model has significant qualitative effects on $P_c$ or $P_c/P_{\rm gas}$. For all four cases shown, the CR pressure ranges from $P_c/k_B \sim 10^{3.5}$ K cm$^{-3}$ in sub-Milky Way galaxies to $\sim 10^7$ K cm$^{-3}$ in the brightest starbursts, while the ratio of CR pressure to gas pressure ranges from $\sim 1$ in sub-MW galaxies to $\sim 10^{-3}$ in starbursts. There are differences at the factor of few level, but nothing larger.

Turning to the third row, we encounter a very different situation. The calorimetric fraction is systematically much lower for the ``Streaming'' than for the ``Scattering'' transport model. The former has calorimetric fractions of $5-10\%$ for Milky Way-like conditions rising to at most $\sim 50\%$ in starbursts, while the latter has calorimetric fractions that are at least $\sim 50\%$ for Milky Way-like galaxies, rising to nearly 100\% in the starburst regime. These differences are also apparent in \autoref{fig:ks_plane_cut_pressure} and \autoref{tab:calorimetry}, where we show results along the \citet{Kennicutt1998} relation, and for a sample of \textit{Fermi}-detected local galaxies, respectively. Clearly the choice of ``Scattering'' or ``Streaming'' leads to significant changes in the degree of calorimetry. The results also depend substantially on the Alfv\'en Mach number: even a factor of two change in this quantity produces noticeable changes in $f_{\rm cal}$.
These changes are in opposite directions and of different sizes for the two possible models, however: increasing $M_A$ lowers $f_{\rm cal}$ for the ``Streaming'' model, while raising it for the ``Scattering'' model. Most of the difference between the ``Streaming'' and ``Scattering'' models can be traced to the comparatively larger value of $\tau_{\rm stream}$ in the streaming model, where the low ion fraction allows fast streaming and thus efficient dissipation. 

Based on our analysis in this section, we can see that our conclusions regarding the typical CR pressure and pressure fraction are robust and depend only very weakly on the transport model we adopt. They are ultimately driven by the fact that, regardless of the transport model, lines of constant $P_c/P_{\rm gas}$ correspond to loci of slope close to 2 in the plane of $\Sigma_{\rm gas}$ versus $\dot{\Sigma}_\star$, while the observed relation between these two quantities is not so steep. Our conclusion that the degree of calorimetry increases from low to high surface density galaxies is similarly robust against the transport model we adopt, but the absolute values of the calorimetric fraction are much less so. These appear to depend sensitively on the exact values of the streaming and absorption optical depths, which are functions of the transport model, and are quite sensitive to parameters such as the Alfv\'en Mach number and (for the scattering model) the CR energy. It is also worth noting that in real galaxies both the ``Streaming'' and ``Scattering'' transport mechanisms likely co-exist: some CRs are deposited in the neutral phase and experience the former, while some enter the ionised phase and experience the latter; there may also be significant exchange of CRs between the phases. The true degree of calorimetry averaged over the galaxy as a whole is therefore likely to be somewhere in between the two limiting cases that we have explored.

\subsection{Caveats and limitations}
\label{ssec:caveats}

Our treatment is a semi-analytic, 1-D study and, as such, it cannot fully capture the complexity of the real world.
In particular, we assume a smoothly evolving ISM density profile while, of course, the real ISMs of galaxies have highly intermittent, multi-phase structures. 
This intermittency has been shown by some of us in previous studies \citep[][also see a number of subsequent works by others]{Krumholz2012,Krumholz2013} to have  important implications for  indirect radiation pressure feedback where, in particular, gas clumpiness renders photons less efficacious in driving global outflows than one would estimate assuming a smooth ISM density.
On the other hand, \citet{Thompson2016} showed, again in the context of radiaiton pressure feedback, that even for a system that is globally ``sub-Eddington'' precisely this ISM intermittency means that photons can launch local outflows from individual, low-surface-density patches of the gas distribution.
Ultimately, absent full numerical studies -- to which we look forward --
in general we cannot be sure about the effect of clumping on CR wind driving.
There is, however, a qualitative argument we can adduce that suggests that ISM intermittency may be a less important effect for CRs than photons:
the case presented by CRs is qualitatively different to photons because the former move along field lines that thread through both dense clumps and low density gas, effectively connecting these different phases. This leads us to the qualitative expectation that CRs should be relatively more confined than photons and, therefore, better coupled to the dense gas because of the magnetic field lines that thread throughout the gas. 
The main exception to this statement will be CRs that are trapped in hot gas that leaves the galaxy at high speed as part of a wind. 
As discussed above, some fraction of the CRs may not interact with the neutral ISM at all, and thus the main effect of advective escape is likely to be an effective reduction in the $\epsilon_{c,1/2}$ parameter that describes the CR energy per unit mass of stars formed that is injected into the neutral ISM.
Note, however,  that
the very fact that $\gamma$-ray emission (that is almost certainly dominantly hadronic)
is detected from a number of nearby starbursts means that there is an implicit limit here: at least some CRs have to interact with neutral gas before escaping.
This consideration was rendered quantitative by \citet{Lacki2010} and \citet{Lacki2011}
who showed that, granted that CRs are energised by star-formation, there are firm, and rather constraining, lower limits on the effective gas density the typical CR ``sees" in escaping a starburst.

\section{Summary and conclusions}
\label{sec:discussion}

In this paper we employ an idealised, slab model of galactic discs to investigate the large-scale, dynamical importance of cosmic rays (CRs) in supporting the neutral, star-forming interstellar medium (ISM) across the full sequence of star-forming galaxies, from near-quiescent dwarfs to intense starbursts. Our ultimate goal is to determine under what conditions we expect CRs to make a substantial contribution to the pressure balance of the ISM, as is the case in the Milky Way, and to what extent the role of CR pressure is correlated with the degree of calorimetry in galaxies, i.e., the fraction of CRs injected into a galaxy that ultimately produce pions and thence $\gamma$-rays.
In our model, the vertical column of gas in a galactic disc is maintained in hydrostatic balance by the competition between stellar and gas self-gravity, and a combination of turbulent and CR pressure. CRs generated near the midplane travel vertically through the gas column, undergoing losses due to both ``absorption'' (i.e., pion-producing $pp$-collisions) and streaming instability as they do so.
We show that this system is characterised primarily by three dimensionless numbers: $\tau_{\rm stream}$, $\tau_{\rm abs}$, and $f_{\rm Edd}$ (as given most generally by 
equations
\ref{eq:tausDefn}, \ref{eq:tauaDefn}, and \ref{BC_3}, respectively).
These parameterise, respectively, the streaming and absorptive optical depths presented by the gas column to the CRs, and the ratio of the CR momentum flux to the gravitational momentum flux, i.e., the CR Eddington ratio.

For any given combination of these three parameters, together with a total gas fraction, we can obtain solutions for the gas density and CR energy density as a function of height, from which we derive our two parameters of interest: the fractional pressure provided by CRs, and the calorimetry fraction, i.e., the fraction of CR flux that is lost to $pp$ collisions, and thus becomes available to produce observable $\gamma$-ray emission. We show that the CR pressure fraction is primarily determined by $f_{\rm Edd}$, and increases with $f_{\rm Edd}$ from small values for $f_{\rm Edd} \ll 1$ to values of order unity for $f_{\rm Edd} \sim 1$, up to a critical value of $f_{\rm Edd}$ beyond which hydrostatic equilibrium is impossible; we discuss the implications of this finding further in \citetalias{Crocker2020b} in this series. By contrast, the degree of calorimetry is controlled primarily by the optical depths, and is insensitive to the Eddington ratio. Calorimetry is maximised when $\tau_{\rm abs} \gg \tau_{\rm stream}$ and $\tau_{\rm abs} \gg 1$.

In order to draw conclusions about observed galaxies, we develop a model to estimate the dimensionless quantities $\tau_{\rm stream}$ (\autoref{eq:tau_s_stream} or \autoref{eq:tau_s_scat}, for ``Streaming'' or ``Scattering'' transport of CRs, respectively), $\tau_{\rm abs}$ (\autoref{eq:tau_s_stream} or \autoref{eq:tau_a_scat}), and $f_{\rm Edd}$ (\autoref{eq:fEdd_scaling}) from observations, primarily the gas and star formation surface densities of galaxies -- the former determines the optical depth and the strength of gravitational confinement (the denominator in $f_{\rm Edd}$), while the latter determines the CR flux per unit area entering the ISM (the numerator in $f_{\rm Edd}$). While these quantities broadly constrain the dimensionless parameters in our model, in detail the mapping between observables and dimensionless quantities depends on the microphysics of CR transport. We therefore consider a range of transport models, corresponding to differing assumptions about the phase of the ISM through which CRs travel, and the mechanism by which they interact with MHD turbulence in the ISM.

Independent of assumptions about transport mode, however, we show that, as the gas column density is dialed upwards, galaxies become increasingly calorimetric 
and are, therefore,
increasingly good
$\gamma$-ray sources (see \autoref{fig:ks_plane_cut_pressure} and \autoref{tab:calorimetry}; cf.~\citealt{Torres2004,Thompson2007,Lacki2010,Lacki2011,Yoast-Hull2016,Peretti2019}).
CRs are never dynamically important on global scales for gas surface densities exceeding $\sim 10^{2.5} \ \msun$ pc$^{-2}$  (\autoref{fig:ks_plane_pressure} and \autoref{fig:ks_plane_crmodels}), and indeed above a gas surface density of $\sim20$\,M$_\odot$ pc$^{-2}$, the pressure declines rapidly (see \autoref{fig:ks_plane_cut_pressure}). In the densest starbursts, the ratio of CR to other pressures drops to only $\sim 10^{-3}-10^{-4}$.
Conversely, at lower surface gas densities CRs can take on considerable dynamical significance, providing pressure comparable to the gas pressure, but at the same time these galaxies are substantially sub-calorimetric. As is implicit in the results of \cite{Jubelgas2008} and as discussed in \cite{Socrates2008}, the ultimate factor driving the trend toward smaller dynamical importance for CRs in more dense and intensely star-forming galaxies is rapid pionic losses. As discussed in the context of the radio and gamma-ray emission from star-forming galaxies across the Kennicutt-Schmidt law by \citet{Lacki2010,Lacki2011} (see also \citealt{Thompson2013}), the distribution of observed galaxies in the plane of gas and star formation surface densities guarantees that high gas surface density systems will have the smallest overall CR pressure relative to that required for hydrostatic equilibrium. At high gas surface densities where pion production is the dominant loss mechanism for CRs, the CR pressure scales with star formation rate and gas surface density as 
\begin{eqnarray}
    P_{\rm CR} &\propto& \frac{\dot{E}_{\rm CR}t_{\rm col}}{\rm Volume}\propto \frac{\rm SFR}{\pi R^2}\frac{t_{\rm col}}{h}\propto \frac{\dot{\Sigma}_\star}{\Sigma_{\rm gas}} \nonumber \\
    &\simeq&1\times10^{5}\,{\rm K\,\,cm^{-3}}\,\dot{\Sigma}_{\star, -1}\,\Sigma_{\rm gas, 2}^{-1}
\end{eqnarray}
where SFR is the total star formation rate, $\dot{E}_{\rm CR}$ is the total CR energy injection rate, $t_{\rm col}$ is the pion loss timescale (eqs.~\ref{eq:tcol} and \ref{eq:tcol2}), and the approximate equality in the second line provides a numerical value scaled to $\dot{\Sigma}_{\star, -1}=\dot{\Sigma}_\star/(0.1\,\rm M_\odot \,\,pc^{-2}\,\,Myr^{-1})$ and $\Sigma_{\rm gas, 2}=\Sigma_{\rm gas}/(100\,\rm M_\odot\,\,pc^{-2})$. By contrast, the self-gravitational pressure of a galactic disc scales as $P_*\propto \Sigma_{\rm gas}^2$, so that $$P_{\rm CR}/P_* \propto \dot{\Sigma}_\star/\Sigma_{\rm gas}^3.$$ Thus, maintaining constant $P_{\rm CR}/P_*$ would require that the star formation rate surface density increase as the cube of the gas surface density. This corresponds to a Kennicutt-Schmidt relation with index of 3, whereas the observed index of the relation is much shallower and ranges from $\approx 1-2$. The decline in the dynamical importance of CRs at high surface density follows directly from this. At the same time, galaxies with higher gas surface densities do have higher absorption optical depths, and thus are more calorimetric. This combination drives the anti-correlation between CR dynamical importance and calorimetry in galaxies \citep{Lacki2011,Thompson2013}.

Finally, we remark that the model we have presented here has obvious further applications. We have already mentioned one of these, which is the subject of \citetalias{Crocker2020b}: launching of CR-driven cool galactic winds. A further follow-up is to combine our results here with those of \citet{Crocker2018}, who develop a similar plane-parallel atmosphere model for radiation transport out of galactic discs. Combining these models will yield fully self-consistent predictions for the run of gas density, CR energy density, and radiation energy, and thus for the synchrotron and inverse Compton emission produced by leptonic CRs. This will be the subject of future work.

\section*{Data Availability Statement}

No new data were generated or analysed in support of this research.

\section*{Acknowledgements}

This research was funded by the Australian Government through the Australian Research Council, awards FT180100375 (MRK) and DP190101258 (RMC, MRK, and TAT). RMC gratefully acknowledges conversations with Felix Aharonian, Geoff Bicknell, Yuval Birnboim, Luke Drury, Alex Lazarian, Chris McKee, Christoph Pfrommer, Heinz V{\"o}lk, and Siyao Xu. MRK and TAT acknowledge support from the Simons Foundation through the Simons Symposium Series ``Galactic Superwinds: Beyond Phenomenology'', during which they collaborated on this work. MRK carried out part of this work during a sabbatical funded by an Alexander von Humboldt Research Award. TAT thanks Brian Lacki, Eliot Quataert, Ben Buckman, and Tim Linden for discussions and collaboration. TAT is supported in part by National Science Foundation Grant \#1516967 and NASA ATP 80NSSC18K0526.
We thank the anonymous referee for a constructive and useful report.








\appendix

\section{On the streaming speed of $\sim$GeV CRs in local spiral and dwarf galaxies}
\label{app:streamSpeed}

In the main text and in \citetalias{Crocker2020b} the form of the
diffusion coefficient we adopt in the case of streaming and
for the dynamically dominant $\sim$GeV CRs
assumes that the
streaming speed is identical to the ion Alfv\'en speed. 
We showed that this is an accurate approximation for starburst environments in \citet{Krumholz2019}.
Here we show that it remains a tolerably accurate assumption down to the much lower
gas densities typical of the neutral ISM in
local spirals and dwarfs.
Moreover, as we also show, the assumption remains valid
irrespective of the identity of the 
dominant ionised species (in particular, whether the dominant species is C$^+$ at $\chi \sim 10^{-4}$ in
starbursts or H$^+$ at $\chi \sim 10^{-2}$ in Milky Way conditions).

First, we recapitulate the results for starburst-like ISM conditions derived in \citet{Krumholz2019}.
Make the empirically-motivated assumption that the CRs fall into a power-law distribution with respect to Lorentz factor $\gamma$, $d\nCR/d\gamma\propto \gamma^{-p}$. 
Then we can write $\nCR(>\gamma) = C \gamma^{-p+1}$; for the Milky Way near the Solar Circle, $C = C_\mathrm{MW} \approx 2\times 10^{-10}$ cm$^{-3}$ and $p\approx 2.6$ \citep{Wentzel1974,Farmer2004}. 
Adopting this functional form for $\nCR(>\gamma)$, \citet{Krumholz2019} show that the growth rate of the streaming instability balances the rate at which it damps due to ion-neutral drag if the CR streaming velocity relative to the ion Alfv\'en speed obeys
\begin{eqnarray}
\frac{\vst}{\vAi}-1 & = & 
\frac{\gd \chi \MA c}{4 C e \uLA \mui \gamma^{-p+1}} \sqrt{\frac{\mH m^2 \muH^3 \nH^3}{\pi}}
\nonumber \\
& \simeq & 2.3\times 10^{-3} \frac{\ECRz^{p-1} \nHthree^{3/2} \chi_{-4}\MA}{C_3 \uLAone},
\label{eq:vratio}
\end{eqnarray}
where $\gamma_{\rm d}$ is the ion-neutral drag coefficient, $\mu_i$ is the mean particle mass of the ions in units of amu, $\uLA$ is the turbulent velocity of Alfv\'{e}nic modes at the injection scale of the turbulence (which we can, with good accuracy, take to be identical to $\sigma$), $C_3 = C/1000 C_\mathrm{MW} = C/2\times 10^{-7}$ cm$^{-3}$, and the numerical evaluation is for CR protons ($m=\mH$) and, in the first instance, a population of ions dominated by C$^+$ ($\mui=12$). 

Note from \autoref{eq:vratio} that, 
while the CR energy density is significantly in excess of Milky Way values, we are guaranteed that the streaming velocity will be very close to $\vAi$, the ion Alfv\'en speed.
For Milky Way conditions, we need to renormalise \autoref{eq:vratio} and we now take the dominant ionised species to be protons ($\mu_i \to 1$) and assume a much lower CR number density $C\sim C_{\rm MW}$ and gas number density $n_{\rm H}\sim 1$ cm$^{-3}$ (the mean for Milky Way midplane) and a much higher ionisation fraction $\chi\sim 10^{-2}$ \citep{Wolfire2003}, in which case we find:
\begin{eqnarray}
\frac{\vst}{\vAi}-1 
& \simeq & 0.086 \frac{\ECRz^{p-1} 
n_{\rm H,0}^{3/2} \chi_{-2}\MA}{C_0 \uLAone},
\label{eq:vratio2}
\end{eqnarray}
where $C_0 \equiv C/C_\mathrm{MW} = C/2\times 10^{-10}$ cm$^{-3}$ and $n_{\rm H,0} \equiv n_{\rm H}/(1$ cm$^{-3})$.
Comparing \autoref{eq:vratio} and \autoref{eq:vratio2},
one can see that, normalising to Milky Way like conditions and, in particular, accounting for the increase in ion neutral drag when going from  C$^+$ to mostly from protons, then the streaming speed goes up from 0.2\% above the Alfv\'en ion speed to $\sim$10\% above the Alfv\'en ion for $\sim$GeV CRs.
For dwarfs, sub-Milky-Way CR energy densities will push the streaming speed to be still larger relative to the ion Alfv\'en speed, but this is somewhat ameliorated by the corresponding decline in 
neutral ISM number density; we estimate that for the most extreme dwarfs in the parameter space one might reach a streaming speed $\sim 50$ \% in excess of $\vAi$.
Altogether, we take these calculations to 
indicate that setting $v_s = \vAi$ for GeV CRs is a 
well-justified assumption over the entire Kennicutt-Schmidt plane.

\section{Parker stability}
\label{app:parker}

In this appendix we consider the possible impact of  \citet{Parker1966} stability on our conclusions. Since Parker's initial calculation, a number of authors have extended the analysis include the effects of CR diffusion and streaming \citep{Ryu2003, Rodrigues2016, Heintz2018, Heintz2020}, which are generally destabilising. The basic conclusion of this analysis is that galactic discs across a wide range of parameter space are generally subject to Parker instability. We therefore expect that many of our models will be Parker unstable. However, this seems unlikely to modify our conclusions substantively, for the following reason: the effect of the instability is to drive turbulent motions, and to increase CR flux compared to our estimates, i.e., to allow CRs to propagate through the gas more rapidly than our laminar calculation suggests. The effect of turbulence is already included empirically in our models, since we simply rely on observed gas velocity dispersions, and thus is not important. In principle, however, the increased flux made possible by the instability could lead to lower CR pressures near the midplane, which would increase the stability of the system against self-gravity.

However, there is an important limit to the amount by which Parker instability might contribute to the flux, which is that the \textit{mechanism} by which Parker instability increases the flux is through convective motions of the gas. When the instability occurs, buoyant magnetic field lines and their associated CRs rise in arches, while gas falls between the arches into valleys. Because the transport is convective, the maximum CR flux that results from Parker instability is ultimately limited by the speed of the rising arches that carry the CRs: $F_{c,\rm P} < \sigma u_c$; the true transport rate is certainly smaller than this, since this is the flux that would apply if the gas were moving uniformly at speed $\sigma$. In terms of our dimensionless variables, we can therefore write the ratio of the Parker instability-driven flux $F_{c,\rm P}$ to the diffusive flux $F_c$ (\autoref{eq:Fc}) as
\begin{equation}
    \frac{F_{c,\rm P}}{F_c} < \frac{3}{K_* f_c},
\end{equation}
where $f_c$ is the dimensionless CR propagation speed defined by \autoref{eq:numerical_eqns}. This ratio takes on its maximum value as $\xi\to\infty$, where it is
\begin{equation}
    \frac{F_{c,\rm P}}{F_c} < \frac{3}{4 K_* \tau_{\rm stream}} = \frac{3\sigma}{4 v_s} = \frac{3}{2\sqrt{2}} M_A \left(\sqrt{\chi},1\right),
\end{equation}
where the first term in parentheses applies for the streaming propagation model, and the second for the scattering or constant $\kappa_*$ models. Thus we see that, unless $\mathcal{M}_A \gg 1$ (and $\gg 1/\sqrt{\chi}$ for the streaming transport model), CR transport via Parker instability is always comparable to or smaller than the flux we have already included in our models. We therefore conclude the Parker instability cannot significantly alter our conclusions.

\bsp	
\label{lastpage} 
\end{document}